\RequirePackage{fix-cm}
\documentclass[smallextended]{svjour3}       
\smartqed  
\usepackage{natbib}
\usepackage{aas_macros}
\usepackage{xcolor}
\usepackage{siunitx}
\usepackage{wasysym}
\usepackage{soul}
\usepackage[finalnew]{trackchanges} 
\addeditor{MW}
\usepackage{graphicx}

\begin{document}

\title{Synergies between Venus \& Exoplanetary Observations
}
\subtitle{Venus and its extrasolar siblings}


\author{M.J. Way \and
        Colby Ostberg \and
        Bradford J. Foley \and
        Cedric Gillmann \and
        Dennis H\"oning \and
        Helmut Lammer \and
        Joseph O'Rourke \and
        Moa Persson \and
        Ana-Catalina Plesa \and
        Arnaud Salvador \and
        Manuel Scherf \and
        Matthew Weller 
}


\institute{M.J. Way \at
              NASA Goddard Institute for Space Studies,
              2880 Broadway, New York, NY 10025, USA\\
              Theoretical Astrophysics,
              Department of Physics and Astronomy,
              Uppsala University, Uppsala, Sweden\\
              \email{Michael.J.Way@nasa.gov}
              \and
           Colby Ostberg \at
           Department of Earth and Planetary Sciences, University of California, Riverside, CA 92521, USA
           \and
           Bradford J. Foley \at
           Department of Geosciences, Pennsylvania State University, University Park, Pennsylvania
           \and
            Cedric Gillmann \at
               Department of Earth, Environmental and Planetary Sciences, Rice University, Houston, TX 77005, USA
           \and
           Dennis H\"oning \at
               Potsdam Institute for Climate Impact Research, Potsdam, Germany\\
               Department of Earth Sciences, Vrije Universiteit Amsterdam, The Netherlands
           \and
           Helmut Lammer \at
               Space Research Institute, Austrian Academy of Sciences, Schmiedlstr. 6, 8042, Graz, Austria
           \and
           Joseph O'Rourke \at
               School of Earth and Space Exploration, Arizona State University, Tempe, AZ, USA
           \and
           Moa Persson \at 
                Institut de Recherche en Astrophysique et Planétologie, Centre National de la Recherche Scientifique, Université Paul Sabatier - Toulouse III, Centre National d’Etudes Spatiales, Toulouse, France
           \and
           Ana-Catalina Plesa \at
               Institute of Planetary Research, DLR, Berlin, Germany\\
               \email{Ana.Plesa@dlr.de}
           \and
           Arnaud Salvador \at
             Department of Astronomy and Planetary Science, Northern Arizona University, Box 6010, Flagstaff, AZ 86011, USA\\
             Habitability, Atmospheres, and Biosignatures Laboratory, University of Arizona, Tucson, AZ, USA\\
             Lunar and Planetary Laboratory, University of Arizona, Tucson, AZ, USA
           \and
           Manuel Scherf \at
               Space Research Institute, Austrian Academy of Sciences, Schmiedlstr. 6, 8042, Graz, Austria\\
               Institute of Physics, University of Graz, Graz, Austria\\
              Institute for Geodesy, Technical University, Graz, Austria
           \and
           Matthew Weller \at
               Lunar and Planetary Institute, 3600 Bay Area Blvd.,Houston, TX 77058
}

\date{Received: 13 April 2022 / Accepted: 11 January 2023}

\maketitle

\begin{abstract}
In this chapter we examine how our knowledge of present day Venus can inform terrestrial exoplanetary science and how exoplanetary science can inform our study of Venus. In a superficial way the contrasts in knowledge appear stark. We have been looking at Venus for millennia and studying it via telescopic observations for centuries. Spacecraft observations began with Mariner 2 in 1962 when we confirmed that Venus was a hothouse planet, rather than the tropical paradise science fiction pictured. As long as our level of exploration and understanding of Venus remains far below that of Mars, major questions will endure. On the other hand, exoplanetary science has grown leaps and bounds since the discovery of Pegasus 51b in 1995, not too long after the golden years of Venus spacecraft missions came to an end with the Magellan Mission in 1994. Multi-million to billion dollar/euro exoplanet focused spacecraft missions such as JWST, and its successors will be flown in the coming decades. At the same time, excitement about Venus exploration is blooming again with a number of confirmed and proposed missions in the coming decades from India, Russia, Japan, the European Space Agency (ESA) and the National Aeronautics and Space Administration (NASA). In this chapter, we review what is known and what we may discover tomorrow in complementary studies of Venus and its exoplanetary cousins.
\end{abstract}

\section{Can exoplanets inform Venus’ evolutionary history?}\label{sec:01b1}
It may sound preposterous to propose that terrestrial exoplanets, which are far from being explored in-situ, and which present challenges even to detection of their atmospheres, can in any way inform Venus’ evolutionary history. Yet exoplanetary science has already provided a means to put ancient Venus 4.2 billion years ago within the habitable zone \citep{Yang2014,Way2016}. Initial studies of Venus’ early climate by \cite{Ingersoll1969,Pollack1971,Kasting1984}, and others laid out the challenges for Venus having temperate surface conditions in its early history, given the $\sim$40\% higher incident solar radiation it received 4.2Ga compared with modern-day Earth. However, \cite{Pollack1971} demonstrated that temperate conditions were possible if Venus had 100\% cloud cover, providing an albedo sufficiently high to block enough incoming sunlight to reduce surface temperatures to less than 300K. Yet he provided no rationale for his choice of 100\% cloud cover. 
Moving 40+ years into the future exoplanet researchers were beginning to look at large parameter sweeps using 3-D General Circulation Models (GCMs) to investigate how insolation and rotation rate influence climate \citep[e.g.][]{Yang2014}. This effort was driven in part by the discovery of a large number of planets orbiting M-dwarf and K-dwarf stars -- many in their habitable zones. One of the first of these exoplanet studies by \cite{Leconte2013} used the Laboratoire de M\'et\'eorologie Dynamique (LMD)\footnote{https://www-planets.lmd.jussieu.fr/} GCM to demonstrate that temperate conditions were possible for the tidally locked world HD 85512 b, which orbits a K-dwarf star with a 58-day period. A year later, using the National Center for Atmospheric Research (NCAR)\footnote{https://ncar.ucar.edu/} Community Atmosphere Model (CAM) GCM, \cite{Yang2014} demonstrated that slowly rotating worlds (not necessarily tidally locked) with modern Earth-like atmospheres could in fact host temperate surface conditions with mean surface temperatures $<$ 300K at stellar insolations approaching 2.5 times what Earth receives today. This was due to large scale contiguous high albedo tropospheric clouds located in the substellar region. These were a byproduct of the extended single-hemisphere-sized Hadley cells from a weakened Coriolis force due to the slower rotation rate. This exoplanet related discovery had confirmed Pollack’s proposed 100\% cloud cover 43 years later. The \cite{Yang2014} work prompted a number of similar studies \citep{Way2016,Way2018} that confirmed the original result with a completely different 3-D GCM known as ROCKE-3D (Resolving Orbital Keys of Earth and Extraterrestrial Environments with Dynamics)\footnote{https://simplex.giss.nasa.gov/gcm/ROCKE-3D/} \citep{Way2017}. This research has had a profound effect on understanding the possible climate history of Venus and Venus-like worlds. Whereas earlier Venus focused studies claimed an early short-lived habitable period was possible \citep{GrinspoonBullock2007}, these exoplanet studies demonstrated that Venus could have had quite long periods of habitability \citep{Way2020}. 

Thus far \change{three}{at least five} different GCMs have produced the cloud-albedo feedback for slowly rotating worlds: ROCKE-3D, NCAR \citep{Yang2014}, the UK Met Office \change{United}{Unified} Model \citep{Walters2011}, 
LMD, and Exocam
\footnote{https://github.com/storyofthewolf/ExoCAM}. While such coherence may appear definitive these model results must be verified with observations of planets within the canonical Venus Zone \citep[e.g.][hereafter VZ]{Kane2014}. At the same time, there is still great uncertainty related to the longevity of the early magma ocean atmospheres (See Section \ref{sec:VenusMagmaOcean}), in the composition of the atmospheres  \citep[e.g.][]{Bower2022} and exactly what role clouds might play \citep{Turbet2021}. Are these atmospheres a mix of CO, CO$_2$, N$_2$, H$_2$O, CH$_4$, or H$_2$, and what sorts of clouds are involved, if any? Here again exoplanetary observations hold the keys to the kingdom, and are the only way to \change{definitely}{definitively} test and refine our models and their underlying physics.

Planetary scientists recognize that the exploration of Venus can inform our understanding of exoplanets, and vice versa as discussed in this chapter. These linkages permeate the new decadal survey released by the United States of America's National Academies \citep{NAS2021} as detailed in the introduction to this topical collection \citep[][this issue]{ORourke2022}. Table \ref{tab:NAS} pulls verbatim excerpts from this new report identifying some of the observations of Venus and exoplanets that scientists consider most important in the near term. We can study Venus as “the exoplanet in our backyard” and obtain measurements, including in situ data, that are not feasible at planets orbiting distant stars. We can also study a statistical sample of Venus-sized exoplanets to explore if a Venus-like evolutionary pathway is typical. These parallel approaches will promote synergies and strengthen ties between these oft-separated scientific communities.

\begin{table}[ht!]
\scriptsize
\caption{Recently, the Planetary Science and Astrobiology Decadal Survey 2023–2032 highlighted many synergies between observations of Venus and exoplanets \citep{NAS2021}. This report prioritized scientific activities that would help answer two key questions: What does Venus teach us about the evolutionary pathways of exoplanets? Is the evolution of Venus typical of Venus-sized exoplanets? Below, we quoted priority questions, strategic research, and supportive activities from Chapter 15 (``Question 12: Exoplanets") that are related to many of the scientific connections between Venus and exoplanets discussed in this chapter and many others in this collection.}
\label{tab:NAS}
\begin{tabular}{|p{1in}|p{3in}|}
\hline
\multicolumn{2}{|l|}{\bf{Priority questions linking Venus and Exoplanets}}\\
\hline
12.1 & Evolution of the Protoplanetary Disk \\
12.3 & Origin of Earth and Inner Solar System Bodies \\
12.4 & Impacts and Dynamics \\
12.5 & Solid Body Interiors and Surfaces \\
12.6 & Atmosphere and Climate Evolution on Solid Bodies \\
12.10 & Dynamic Habitability \\
12.11 & Search for Life Elsewhere \\
\hline
\multicolumn{2}{|l|}{\bf{Strategic Research to Benefit Exoplanetary Science}}\\
\hline
Question(s)       &  Strategic Research \\
\hline
12.1, 12.3, 12.6  & Measure abundances and isotopic compositions of noble gases and other key elements (in the atmosphere of Venus) \\
\hline
12.6              & Determine the properties of the atmospheres of terrestrial planets (…Venus…) that would be observable on exoplanets \\
\hline
12.10             & Constrain the inner edge of the habitable zone in the solar
system by studying the surface geomorphology and geochemistry of Venus to assess whether it ever possessed oceans \\
\hline
12.11             & Study methods to discriminate past and present false positive
biosignatures on solar system bodies (e.g., abiotic O2 on Venus…) from true biosignatures
to inform false positives discrimination methods for exoplanets\\
                  & \\
                  & Devise metrics and frameworks to establish confidence in interpretation of
biosignatures in the solar system and exoplanetary systems\\
\hline
\multicolumn{2}{|l|}{\bf{Strategic Research on Exoplanets to Benefit Venusian Science}}\\
\hline
Question(s)       &  Strategic Research \\
\hline
12.1              & Characterize protoplanetary disks around young stars\\
\hline
12.3, 12.4, 12.5, 12.6, 12.10 & Obtain an inventory of properties of solid body
exoplanets (i.e., mass, composition, bulk Obtain an inventory of properties of
solid body exoplanets (i.e., mass, composition, bulk atmospheric chemistry and
abundance of clouds and hazes, potential biosignatures, rotation rates, relative
distance from host star, type of host star)\\
\hline
12.4              & Determine how impacts contribute volatiles to (or, in some
cases, remove volatiles from) planetary bodies\\
\hline
12.5              & Search for magnetospheric activity at exoplanets\\
\hline
\multicolumn{2}{|p{4in}|}{\bf{Supportive Activities to Promote Synergy Between Venusian and Exoplanetary Science}}\\
\hline
\multicolumn{2}{|p{4in}|}{Observations of [Venus] through transit spectroscopy and direct-imaging as analogs to exoplanet observations}\\
\multicolumn{2}{|p{4in}|}{}\\
\multicolumn{2}{|p{4in}|}{Observations of particle and gas opacity in [Venus] as
a function of phase angle to help determine the dependence of reflectivity and scattering on particles and clouds}\\
\multicolumn{2}{|p{4in}|}{}\\
\multicolumn{2}{|p{4in}|}{Laboratory studies to understand the relationship between
the bulk composition of a planet and its atmosphere, and to determine the optical properties of clouds and hazes}\\
\multicolumn{2}{|p{4in}|}{}\\
\multicolumn{2}{|p{4in}|}{Increased interactions between the astronomy, planetary science, astrobiology communities}\\
\hline
\end{tabular}
\newline
\vspace{0.3cm}
\newline
\end{table}

\subsection{Transiting Exoplanets in the Venus Zone and JWST}\label{section:01b2}

The Transiting Exoplanet Survey Satellite \citep[TESS;][]{ricker2010transiting} is currently observing our nearest and brightest stellar neighbors in search of exoplanets. Similar to the Kepler/K2 mission \citep[e.g.][and references within]{Howell2014}\footnote{https://www.nasa.gov/mission\_pages/kepler/overview/index.html}, TESS is discovering exoplanets using the transit method. This method works by observing changes in the brightness of a star as a planet passes between the instrument and the star. The magnitude of the change in the star’s brightness reveals the radius of the planet (assuming that one knows the radius of the star), while the periodicity of the brightness fluctuations is used to infer the planet's orbital period. The transit method is intrinsically biased towards planets with shorter orbital periods \citep{kane2008constraining}, since the probability of observing a planet transit is inversely proportional to the planet's orbital period. This observational bias has led to TESS discovering a \change{vast}{large} number of terrestrial planets in the Venus Zone \citep[VZ;][]{Kane2014}. The VZ is defined as the area around a star where a planet is more likely to resemble a Venus analog than an Earth analog, but does not guarantee a planet will have Venus-like surface conditions. Temperate planets may also reside in the VZ, as recent works have highlighted the possibility of Venus sustaining a temperate climate in the past \citep{Way2016, Way2020}. Ultimately, the VZ is a tool to guide target selection for follow-up observations of exoplanet atmospheres. These observations will provide information about the atmospheres of VZ planets, which helps infer information about their surface conditions and test the hypothesis of the VZ. Similar to the Habitable Zone \citep[HZ;][]{Kopparapu2013}, the VZ is defined by two boundaries. The inner VZ boundary is defined, in terms of insolation flux, as 25x the flux received by Earth. This specific value was chosen as it is the flux needed to place Venus on the `Cosmic Shoreline' \citep{zahnle2017cosmic}, which is an empirical relationship used to predict the insolation flux needed for a terrestrial body to lose the majority of its atmosphere via thermal escape processes. The outer VZ boundary is the runaway greenhouse boundary, which is the inner boundary of the HZ. This boundary is the insolation flux where an Earth-like planet is predicted to \change{be forced into}{enter} a runaway greenhouse state.

\begin{figure}
    \centering
    \includegraphics[width=\textwidth]{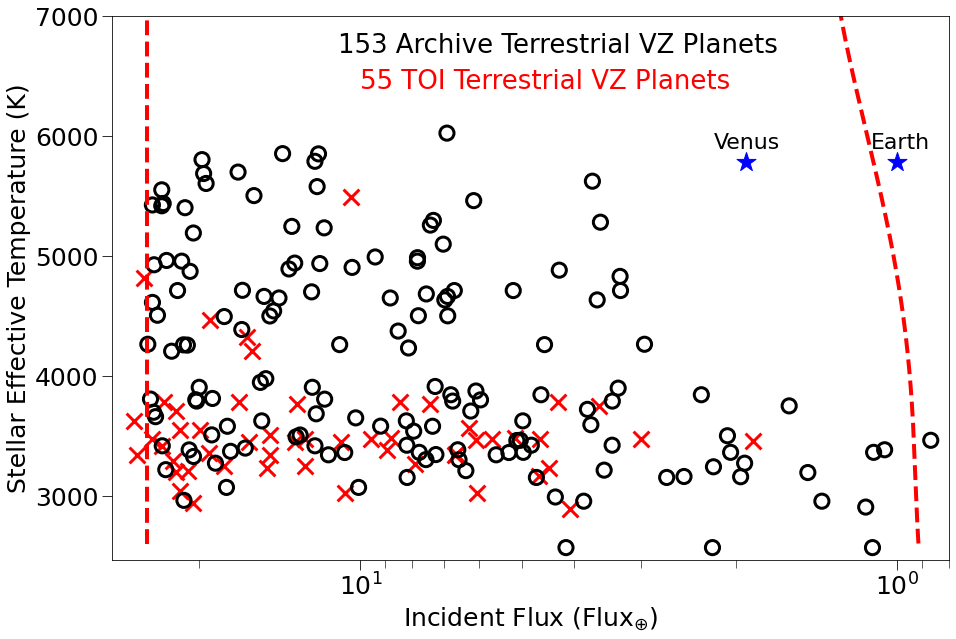}
    \caption{The locations of terrestrial VZ planets (R$_p$ $<$ 1.5R$_\oplus$) from the NASA Exoplanet Archive and TOI list in reference to the VZ as a function of planetary insolation flux. Earth and Venus are shown for reference.}
    \label{fig:TOI_VZ}
\end{figure}

Unlike the Kepler/K2 mission, which observed stars nearly 1000 pc away, TESS is observing stars which are at a distance of $\sim$60 pc. The closer vicinity of TESS stars makes them inherently brighter than Kepler/K2 stars, and therefore allows for more signal to be obtained from them. The increased number of photons from TESS stars creates an excellent opportunity to conduct follow-up observations of the atmospheres of TESS planets from ground and space based instruments. Planets detected by TESS are initially added to the TESS Object of Interest (TOI) list. However a TOI is required to be detected by additional observations in order for it to become a confirmed planet. All confirmed planets are listed on the NASA Exoplanet Archive\footnote{https://exoplanetarchive.ipac.caltech.edu/}. At the time of writing, the NASA Exoplanet Archive and TOI list contain 153 and 55 terrestrial planets (R$_p$ $<$ 1.5R$_\oplus$) that spend any portion of their orbit in the VZ, respectively (Figure \ref{fig:TOI_VZ}). \change{The}{A} radius cutoff of 1.6 R$_\oplus$ \change{was}{is typically} chosen as it is the empirical upper size limit of terrestrial exoplanets \citep{fulton2017}.

Determining that a planet resides in the VZ provides only a first-order estimate about the potential environment on that planet. In order to more accurately deduce possible surface conditions on a VZ planet, observations of its atmosphere will be required. JWST (launched in December 2021) may be humanity's first opportunity to peer into the atmospheres of terrestrial exoplanets via either transmission or secondary eclipse spectroscopy \citep[e.g.][]{barstow2015transit, batalha2017information, beichman2014observations, belu2011primary, clampin2011overview, crouzet2017follow, deming2009discovery, greene2016characterizing, howe2017information, molliere2017observing,Lustig-Yaeger2019,Fauchez2019,Koll2019,Wunderlich2019}.

\subsection{Transmission and Secondary Eclipse Spectroscopy with JWST}
\label{Transmission_Emission_JWST}

Informed predictions of the surface conditions and climates on potential exo-Venuses will require observations of their atmospheres via transmission and secondary eclipse spectroscopy.
Secondary eclipse spectroscopy is conducted by observing the appearance and disappearance of light reflected and/or emitted by the planet as it orbits its host star -- there is no need to spatially resolve the light from the planet from that of the host star. Transmission spectroscopy involves observing starlight that passes through the atmosphere of a transiting exoplanet. Both techniques can be used to gather information about the composition and structure of an exoplanet atmosphere. The atmospheres of terrestrial exoplanets have been inaccessible to this point, but JWST may provide the light-gathering power necessary to retrieve information from terrestrial exoplanet atmospheres \citep[e.g.][]{Lustig-Yaeger2019, batalha2018strategies, morley2017observing, lincowski2019observing,Fauchez2019,Turbet2016,meadows2018habitability}. 

\begin{figure}
    \centering
    \includegraphics[width=\textwidth]{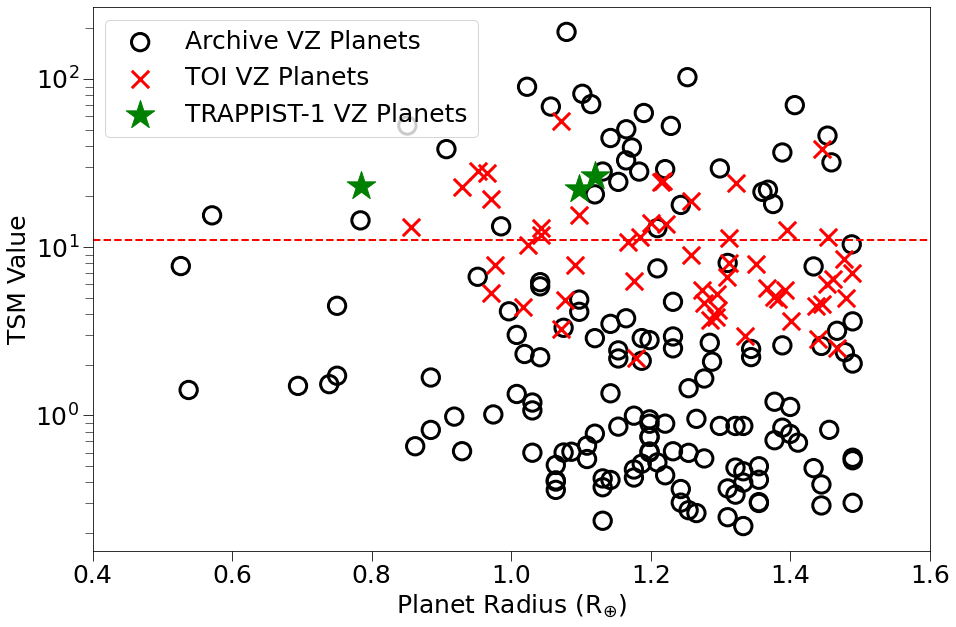}
    \caption{Planetary radii versus associated TSM values for terrestrial planets (R$_p$ $<$ 1.5R$_\oplus$) from the NASA Exoplanet Archive and TOI list. Planets with TSM values greater than 12 (red dotted line) are predicted to allow for a S/N of at least 12 from 10 hours of observations with JWST. The green stars denote the three TRAPPIST-1 planets in the VZ.}
    \label{fig:VZ_TSM}
\end{figure}

The performance of JWST when observing exoplanets can be predicted using the Transmission Spectroscopy Metric \citep[TSM;][]{kempton2018framework}. The TSM provides a first-order approximation of the signal-to-noise ratio (S/N) of transmission spectra resolved from 10 hours of transit observations using the JWST NIRISS instrument \citep{Louie2018} that can be used to prioritize targets that offer the best opportunity for JWST follow-up observations. \cite{kempton2018framework} identified the top terrestrial targets as having TSM values greater than 12.
Applying this threshold to known VZ planets shows there are 36 planets which qualify as top candidates for JWST observations (Figure \ref{fig:VZ_TSM}), including TRAPPIST-1b, c, and d (red stars in Figure \ref{fig:VZ_TSM}). Given that the TRAPPIST-1 system also has 3 planets in the HZ, observations of both the TRAPPIST-1 VZ and HZ planets could help us to discern whether the differences in climate between Earth and Venus is a common phenomena.

\begin{figure}
    \centering
    \includegraphics[width=\textwidth]{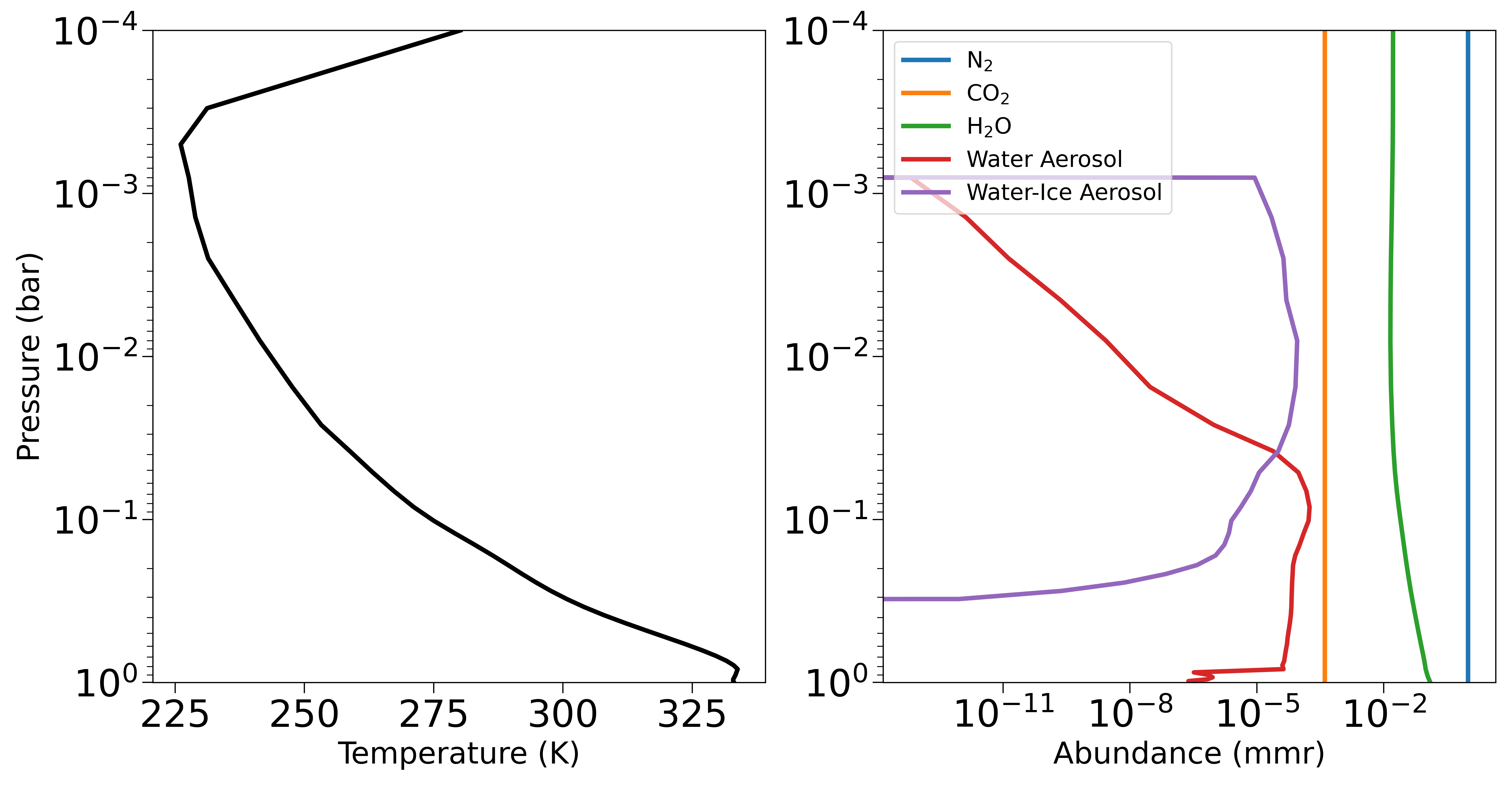}
    \caption{Left: The globally averaged pressure-temperature profile of a Kepler-1649b Exo-Venus hypothetical atmosphere using data from a ROCKE-3D simulation of the planet. Right: Globally averaged Mean Mixing Ratio (mmr) composition versus Pressure. Note that the insolation for this exoplanet has been artificially reduced by a factor of 1.4, otherwise it would have most certainly entered a runaway greenhouse condition.}
    \label{fig:ExoVenus_TP_Comp}
\end{figure}

Here we simulate JWST observations of Kepler-1649b \citep{Angelo2017} as an exo-Venus by modelling hypothetical JWST NIRSpec PRISM transmission spectra using the Planetary Spectrum Generator \citep[PSG; ][]{villanueva2018planetary}. NIRSpec PRISM has a wavelength range of 0.7--5.0 $\mu$m encompassing major H$_2$O and CO$_2$ features, and has been shown to be the optimal instrument for performing transmission spectroscopy in the NIR \citep{Lustig-Yaeger2019}. PSG is a publicly available online interface that couples radiative transfer models, planetary databases, and spectral databases. Exo-Venus transmission or emission spectra can be produced with PSG by superimposing an atmosphere onto a terrestrial exoplanet in the VZ. Kepler-1649b is used as the hypothetical exo-Venus, as its size is similar to that of Venus, with a radius of 1.077 R$_{\venus}$ (1.017 R$_\oplus$), and has a incident insolation flux that is 2.21 times greater than that of Earth (Venus is 1.9), albeit orbiting a much redder M-dwarf star \citep{Angelo2017}. We used an atmosphere for the Kepler-1649b exo-Venus that uses data from a ROCKE-3D simulation of the planet documented in \citet{Kane2018}. Specifically, we use data from simulation 10 in the previously mentioned work, which assumes an Earth-like input atmosphere (1 bar N$_2$ dominated with 376 ppmv CO$_2$), a lower insolation flux than Kepler 1649b of 1.4 and a mean surface temperature of 60$^{\circ}$C making it representative of a \add{hypothetical} temperate ancient-Venus. Note that using the actual insolation flux results in mean surface temperatures well over 100$^{\circ}$C as shown in simulations 1--3 in \citet{Kane2018} \add{which is beyond the capabilities of the GCM used in this study (ROCKE-3D)}. Figure \ref{fig:ExoVenus_TP_Comp} illustrates the structure and chemical composition of the atmosphere from simulation 10.

\begin{figure}
    \centering
    \includegraphics[width=\textwidth]{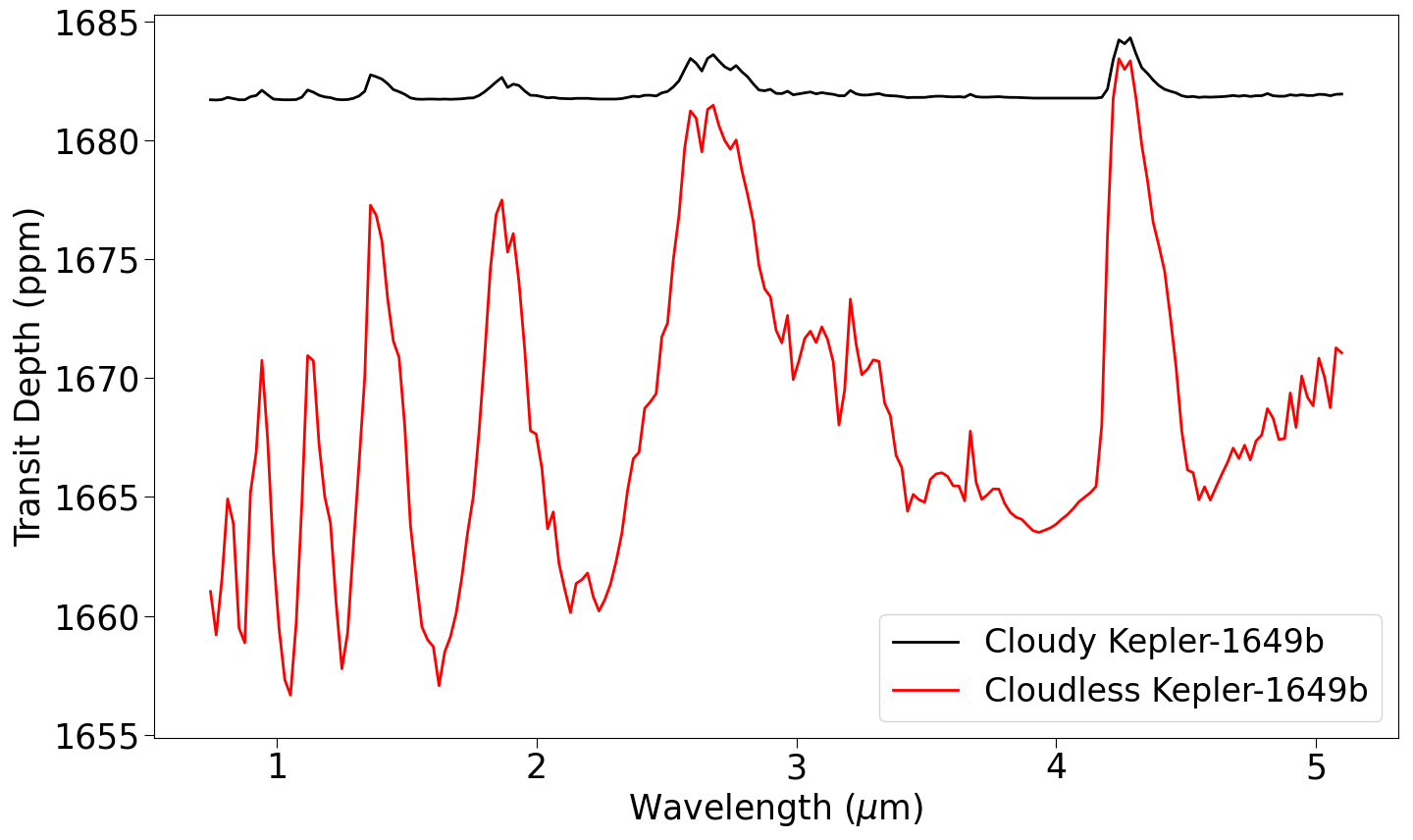}
    \caption{Transmission spectra modelled with PSG for a temperate Kepler-1649b exo-Venus, assuming both a cloudy and cloudless atmosphere.}
    \label{fig:PSG_Venus}
\end{figure}

Using the Kepler-1649b atmosphere from the ROCKE-3D simulation as an input for PSG, we modelled the transmission spectrum of Kepler-1649b from 0.6--5.3 $\mu$m, coinciding with the wavelength range of JWST NIRSpec PRISM. Since PSG is a 1-D radiative transfer model, the globally averaged pressure, temperature, and composition of the simulated Kepler-1649b atmosphere was used. Figure \ref{fig:PSG_Venus} displays the transmission spectra of the Kepler-1649b exo-Venus \add{with and without water and water-ice aerosols, which is hereafter referred to as cloudy and cloudless, respectively.} PSG determined that the atmosphere is opaque at elevations \change{where the clouds were sufficiently dense,}{with higher aerosol densities,} which had a significant affect on the absorption features in the transmission spectra. Prominent H$_2$O and CO$_2$ absorption features are visible in the cloudless spectrum, but are nearly completely truncated by the clouds in the modelled spectrum. The effect of clouds in the temperate Venus atmosphere will likely make it difficult for JWST to detect any absorption features, as shown in previous work \citep{Fauchez2019}. 

The H$_2$SO$_4$ clouds in the atmosphere \change{in}{of} present-day Venus have an equally significant effect on its transmission spectra \citep{ehrenreich2012transmission}. 
This was also demonstrated in \cite{meadows2018habitability} who simulated H$_2$SO$_4$ clouds and hazes in hypothetical modern Venus analogs. Hazes can form when the CH$_4$ to CO$_2$ ratio is greater than 0.1 and are an important contributor to the radiation budget and the detectability of Earth-like planets \citep{Arney2016,Arney2017}. Furthermore, \cite{meadows2018habitability} examined cloud and haze formation effects on the detectability of atmospheres on Proxima Centauri b using a ``1-D coupled climate-photochemical models to generate self-consistent atmospheres for several evolutionary scenarios, including high-O$_2$, high-CO$_2$, and more Earth-like atmospheres, with both oxic and anoxic compositions." They also included the hydrocarbon hazes in instances when the CH$_4$/CO$_2$ ratio was greater than 0.1. Because their atmospheres were not cold enough they did not see any CO$_2$ clouds, but they have been shown to play an important role in the radiation budget in ancient Mars simulations \citep{Colaprete_Toon2003,Forget2013}.
However, it has long been postulated that the H$_2$SO$_4$ clouds on Venus are impermanent and require a regular supply of SO$_2$ from volcanism. As discussed in Section \ref{Volcanism_and_Outgassing} the equilibrium level of SO$_2$ in the atmosphere is set by the volcanic outgassing rate versus the chemical reactions with surface materials \citep{Zolotov2018}. The rate of present day volcanism on Venus is poorly constrained, although there are a number of studies from Venus Express demonstrating hot-spot volcanism \citep{Shalygin2015,Smrekar2010}. Other studies imply geologically recent volcanism due to the radar-dark floors of craters, presumably from volcanic fill-in \cite[e.g.][]{Herrick2011} while others have demonstrated on-going plume activity \citep{Gulcher2020}. Recently, \cite{Byrne2022} have used the recent Earth volcanic record as a proxy to derive estimates for Venus. If volcanism ceased today estimates of the lifetime of the clouds in different studies have ranged from $\sim$2--50 Myr \citep{Fegley1989,Bullock1996,Bullock2001} depending upon surface chemical reaction rates as mentioned above. Hence for some exo-Venus worlds H$_2$SO$_4$ clouds may not be an inhibitor to detection of major atmospheric species for a modern Venus-like atmosphere during \change{low periods of}{periods of low} volcanic sulfur outgassing.

It is important to note that the true nature and variety of environments on Venus-like worlds may be expansive, but will need to be investigated through atmospheric observations of exo-Venus candidates. Additionally, the atmospheric composition of an exo-Venus orbiting an M-dwarf star may differ from that of Venus. Placing Earth around Proxima-Centauri could enhance the \add{abiotic} production of CH$_4$ in its atmosphere \citep{meadows2018habitability} which \change{may be an}{is often cited as an} atmospheric biosignature \citep{Thompson2022}, and the atmospheric composition of Venus may be affected in a similar scenario. Furthermore, from an evolutionary point of view, the large energy deposition from stellar-winds produced by an M-dwarf could, over time, strip molecules from an exo-Venus atmosphere, which would affect the atmospheric composition as well \citep[e.g.][]{Airapetian2020}, but was not accounted for when modelling the Kepler-1649b atmosphere.

\begin{figure}
    \centering
    \includegraphics[width=\textwidth]{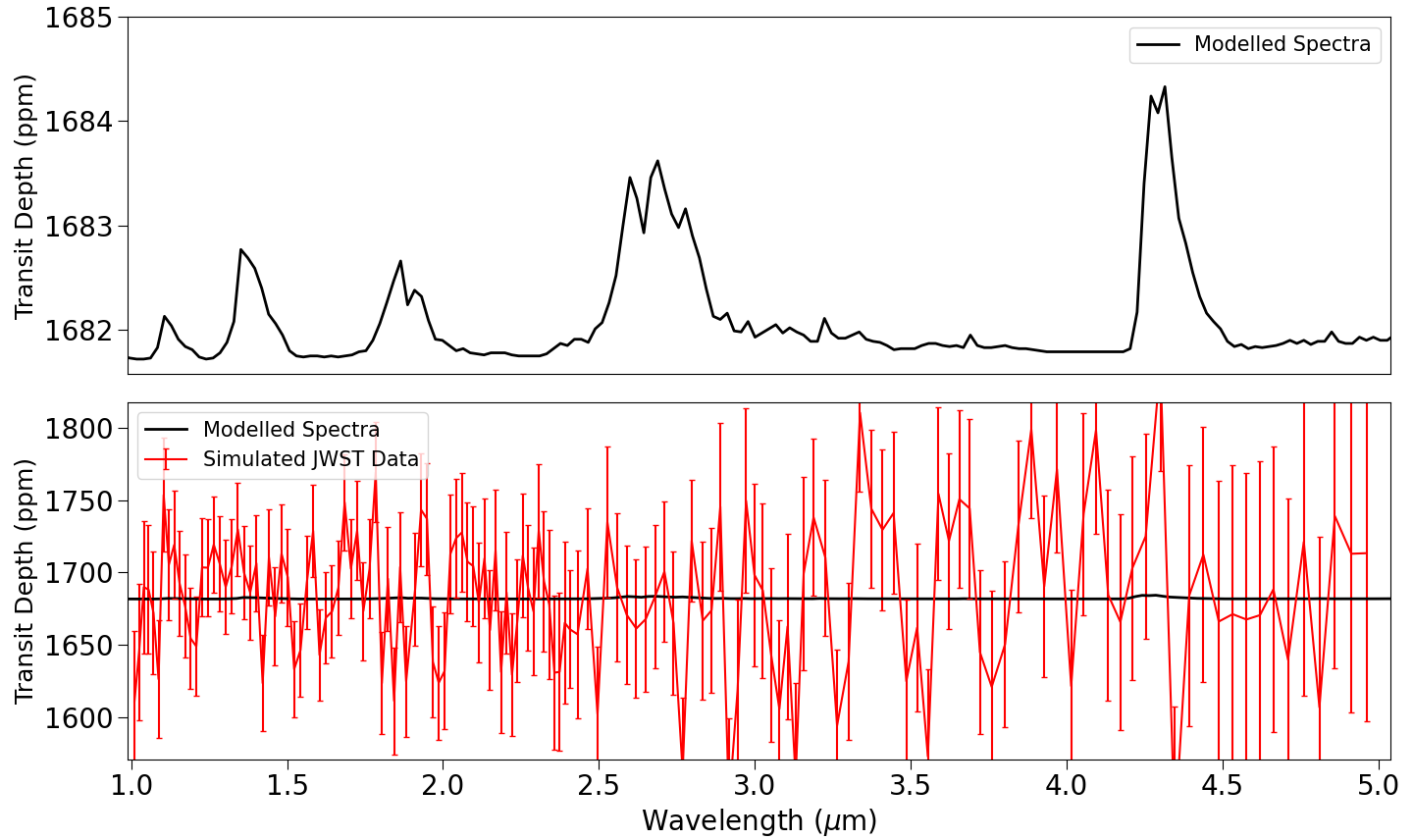}
    \caption{PandExo simulated transmission spectrum of an exo-Venus Kepler-1649b from 30 transit observations using JWST NIRSpec PRISM. The upper figure displays the PSG modelled transmission spectrum with no noise, while the bottom figure compares data from JWST simulated observations of Kepler-1649b to that of the original spectrum. Note that the y-axes of the two plots are on different scales, illustrating the size of the uncertainties in comparison to the noise-less spectrum.}
    \label{fig:Pandexo_Venus}
\end{figure}

The successful detection of \add{transiting} exo-Venus atmospheres with JWST remains uncertain, but models such as PandExo \citep{Batalha2017} can provide insight into how JWST may perform. PandExo is an open-source code that allows users to simulate observations of exoplanets with JWST, and uses the Space Telescope Science Institute's Exposure Time Calculator, Pandeia \citep{Pickering2016}, to predict the S/N of observations. The performance of PandExo's simulated noise has been tested against noise simulations designed by the JWST instrument teams, and is within 10\% agreement of their results \citep{Batalha2017}. Figure \ref{fig:Pandexo_Venus} shows a simulated transmission spectrum of the Kepler-1649b exo-Venus generated by PandExo, assuming 30 transit observations with JWST NIRSpec PRISM. The atmosphere used for the Pandexo simulated observations is the same as that used for Figure \ref{fig:PSG_Venus}. Given 30 transit observation of Kepler-1649b, the simulated JWST data is unable to resolve any of \add{the} major absorption features in the NIR. Furthermore, the large uncertainty in the data would make it difficult to differentiate the spectra from that of a flat-line, which may result in mistaking an exo-Venus as a planet with no atmosphere \citep{Lustig-Yaeger2019b}. Increasing the number of transit observations would decrease the uncertainty in the data, however acquiring the JWST time needed to conduct these observations will be a challenge. 
The features being less than 5 ppm make them smaller than the predicted 20 ppm noise floor of the NIRSpec instrument \citep{rustamkulov2022analysis}, making them potentially undetectable by JWST given any amount of observations and only accessible with future observatories.

\change{Given}{Assuming} that absorption features are detected in the atmosphere of an exoplanet, retrieval algorithms will then be used to estimate its atmospheric composition. Retrieval algorithms have been shown to experience difficulty differentiating Earth-like from Venus-like planets, since Venus' transmission spectra lacks unique absorption features that can be used to distinguish it from Earth \citep{barstow2016telling}. The information gained from a retrieval model can then be applied to a GCM, which model the possible surface conditions of the planet based on the atmosphere estimated by the retrieval. The use of GCMs \change{will be the primary method of}{may play a critical role in} constraining the potential climates of exoplanets \citep{Turbet2016,wolf2019importance} for the foreseeable future \change{given the limitations of}{in coordination with} JWST. 

\begin{figure}
    \centering
    \includegraphics[width=\textwidth]{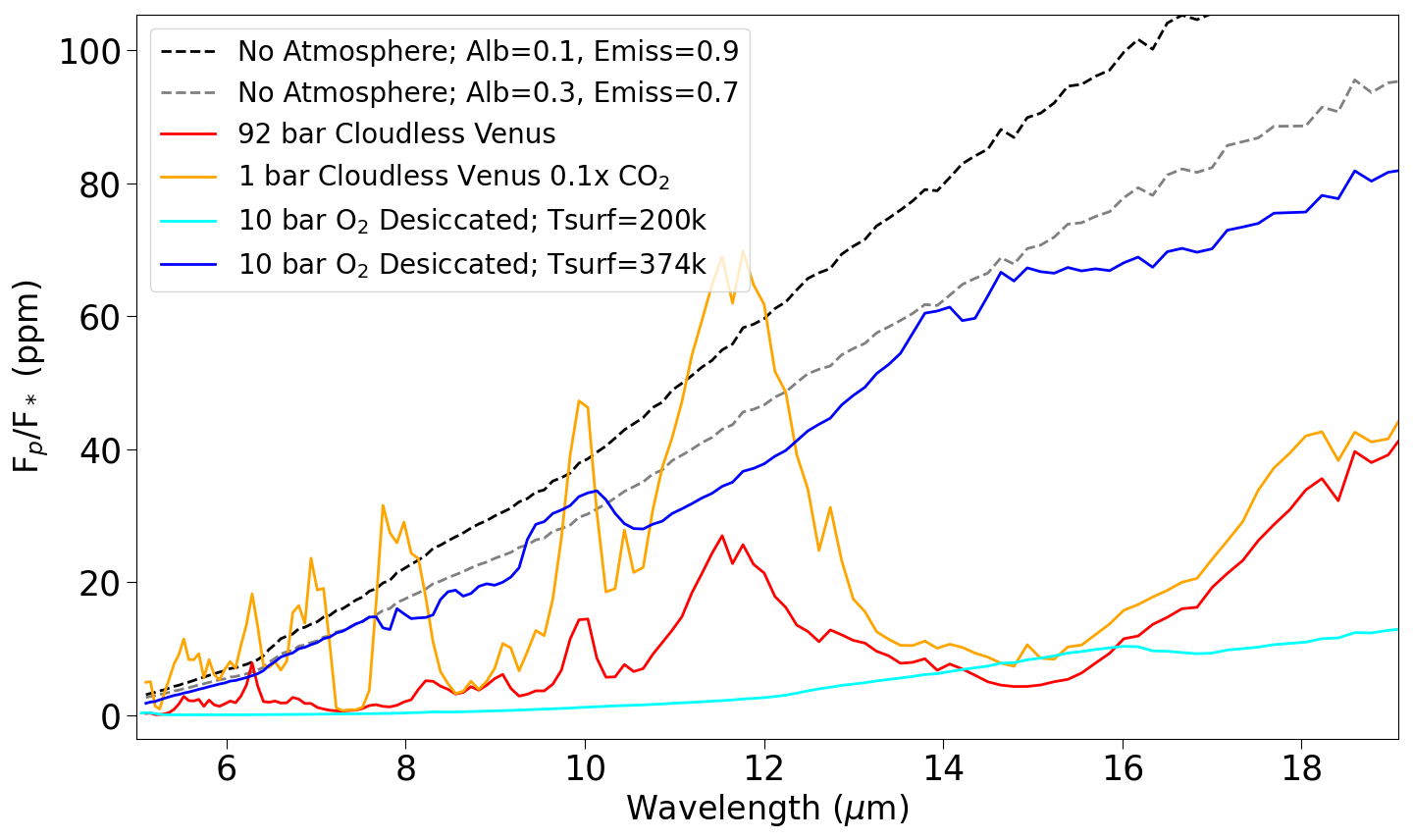}
    \caption{A variety of emission spectra that could be potentially observed on exoplanets using the MIRI instrument aboard JWST. The planet-star flux ratio values are obtained by placing these atmospheres on the Venus-zone planet, L98-59d.}
    \label{fig:Emission_Spectra}
\end{figure}

Emission spectroscopy will be \change{conducted}{attempted} by JWST primarily using the Mid-Infrared Instrument (MIRI), which has a wavelength range between 5 -- 29 $\mu$m. The emission spectra  retrieved by MIRI will be useful for identifying the presence, or lack of an atmosphere on a planet 
\citep{batalha2018strategies,meadows2018habitability,Turbet2016}.
Figure \ref{fig:Emission_Spectra} illustrates several hypothetical emission spectra that could be observed on the VZ planet, L98-59d. Included are the following atmospheres: cloudless 92 bar Venus analog (red); 1 bar cloudless Venus with 0.1$\times$ the CO$_2$ of present-day Venus (yellow) ; 10 bar, O$_2$ dominated desiccated atmosphere with a surface temperature of 374 K; 10 bar, O$_2$ desiccated atmosphere with a surface temperature of 200 K; an atmosphere-less, black-body emission spectrum assuming bond albedo = 0.1 and emissivity = 0.9; an atmosphere-less, black-body emission spectrum assuming a bond albedo = 0.3 and emissivity = 0.7. All atmospheres assume no clouds to illustrate the dependence of emission spectra on atmospheric composition. It can be seen that the presence of CO$_2$ in the 2 Venus-like atmospheres causes the structure of their emission spectra to differ greatly from the other 4 spectra, particularly with the large CO$_2$ emission peaks at 10 and $\sim$12 $\mu$m. The O$_2$ dominated desiccated atmospheres are included since many VZ planets orbit hyperactive M-dwarf stars, which could photodissociate any atmospheric H$_2$O in these planets over time \citep{Wordsworth2013,Luger2015}. In this scenario rapid hydrogen escape would ensue and an O$_2$ dominated, but H$_2$O desiccated, atmosphere would remain. 

\begin{figure}
    \centering
    \includegraphics[width=\textwidth]{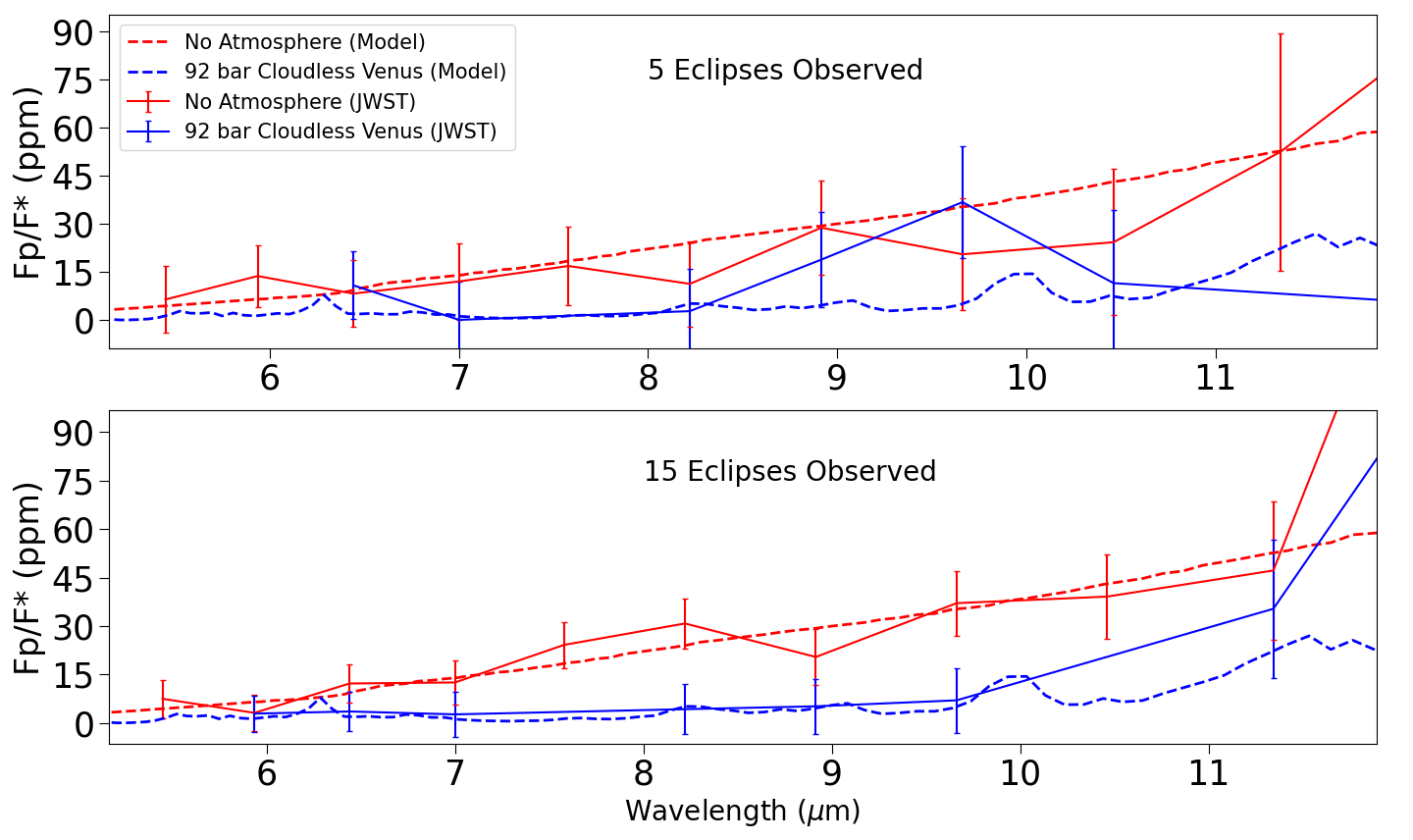}
    \caption{Simulated JWST MIRI LRS data from 5 (top) and 30 (bottom) secondary eclipse observations of L98-59d assuming it has either no atmosphere, or a cloudless 92 bar Venus-like atmosphere. The dotted lines are the PSG modelled emission spectra, while the solid lines are PandExo simulated MIRI observations.}
    \label{fig:PandexoEmission}
\end{figure}

Coupling the PSG emission spectra with PandExo gives insight into the ability of JWST to detect an atmosphere on a hypothetical L98-59d, and whether JWST would be able to tell them apart (Figure \ref{fig:PandexoEmission}). Figure \ref{fig:PandexoEmission} displays simulated JWST data assuming both 5 and 15 secondary eclipse observations of an exo-Venus L98-59d with no atmosphere, and with a cloudless 92 bar Venus-like atmosphere. For 5 eclipse observations, the uncertainty in the simulated data for both cases make it difficult to determine whether there is an atmosphere. With 15 eclipse observations, the simulated data is a much better fit to the modelled spectra up to 11 $\mu$m. Retrieval models will also be used for JWST emission spectra to determine the likelihood of a planet having an atmosphere\add{, but as earlier studies cited above have shown it is unlikely any individual atmospheric features will be discerned.}

In summary, there are an abundance of VZ planets which are promising candidates for follow-up JWST observations, and the TESS mission will be discovering additional candidates throughout its lifetime. Of these candidates, the TRAPPIST-1 planets in the VZ are especially intriguing, as observations of their atmospheres, and the atmospheres of the TRAPPIST-1 HZ planets, will provide an opportunity to compare the differences between Earth and Venus to planets receiving similar insolation flux. JWST will be our first opportunity to obtain information about the atmospheres of terrestrial planets, including exo-Venuses. \add{Simulated JWST data revealed that 15 transit observations with JWST NIRSpec PRISM would be insufficient for resolving the atmosphere of Kepler-1649b with both a temperate exo-Venus, and present-day Venus atmosphere.} Venusian clouds and hazes severely truncate the absorption features in the present-day Venus spectrum, and will make it difficult to efficiently determine the atmospheric composition of an exo-Venus, or detect its atmosphere at all. The temperate exo-Venus atmosphere would be difficult to detect as well, despite the lack of Venus-like clouds. Even if significant JWST time is allotted for observations of exo-Venuses, it still may be the case that atmospheric information vital for understanding the climates of exo-Venuses may remain inaccessible during the JWST era. The inability to infer the surface conditions of exo-Venuses will inhibit exoplanets from being a resource to study Venus' evolution, and whether Venus could have sustained temperate surface conditions in its past. 

\subsection{Future Space and Ground based exo-Venus observational capabilities}\label{sec:Future_Space_Ground}

There are at least three next generation ground-based ($>$20m in diameter) optical near-IR observatories currently under construction (circa 2022) or likely to be built in the near future. The European led Extremely Large Telescope (ELT) has a capable first generation set of instruments \citep{Ramsay2020} and is the only next generation telescope both fully funded and under construction. The Magellan Giant Telescope (GMT) \citep{Fanson2020} and the Thirty Meter Telescope (TMT) \citep{Sanders2013} are yet to be fully funded. The former two are currently under construction in Chile while the TMT is proposed for the northern hemisphere, although the exact location remains uncertain \citep{Clery2019}. Once complete, these new observatories will offer the opportunity for a marked increase in collecting area and resolution. With increasing advances in adaptive optics, they will afford new opportunities to \change{explore}{characterize} the atmospheres of nearby exo-Venuses, as they are discovered by space observatories devoted to detecting such systems via the transit method (e.g. Kepler\footnote{https://www.nasa.gov/mission\_pages/kepler/main/index.html}, TESS\footnote{https://www.nasa.gov/tess-transiting-exoplanet-survey-satellite}, CHEOPS\footnote{https://sci.esa.int/web/cheops}, PLATO\footnote{https://platomission.com/2018/05/07/habitability-of-planets-around-solar-like-stars/}) 
complimented by ground based radial velocity instruments like that of the FLAMES facility at the VLT \citep[e.g.][]{Pasquini2002}.
In space, JWST has just launched. It may be able to detect atmospheres around a few nearby terrestrial planets in systems such as Trappist-1, although such observations will be challenging, as discussed above. 

A mostly-US funded successor to The Hubble Space Telescope was recently recommended as a top priority in the US National Academy of Sciences (NAS) Decadal Survey \cite[][Section 7.4]{NAS2021}\footnote{https://www.nationalacademies.org/our-work/decadal-survey-on-astronomy-and-astrophysics-2020-astro2020}. It is referred to as the ``IR/O/UV Large Strategic Mission" (which we refer to as IROV, see Section 7.5.2 in the NAS report) and recently dubbed the Habitable Worlds Observatory \citep{Clery2023}. It is ``optimized for observing
habitable exoplanets and general astrophysics", according to the report. The UV component is why IROV is more properly termed a successor to The Hubble Space Telescope rather than JWST -- the latter being IR optimized. IROV is scheduled to launch in the early 2040s. IROV is expected to be some combination of The Large UV Optical Infrared Surveyor (LUVOIR) \citep{LUVOIR2019} with 8m diameter and HabEx \citep{Martin2019} with a $\sim$4m diameter mirror, while including a coronagraph for direct imaging and spectroscopy of extrasolar planets. IROV would have a ``light collecting area several times larger, 2-3 times sharper image quality, and instruments and detectors significantly more sensitive, providing 1-2 order-of-magnitude leaps in
sensitivity and performance over HST." The report recommends a $\sim$6m sized mirror as a balance between a Habex 4m, which would struggle to provide a ``robust exoplanet census", and a LUVOIR 8m, which would likely launch much later than IROV, in the late 2040s or early 2050s. As shown in the work of \cite{Checlair2020}, the diameter of the mirror appears to be the critical factor in determining whether we will make the revolutionary discoveries intended. IROV will be capable of observing over 100 nearby Sun-like stars and would quantify the elements of any associated planetary systems, giving ample opportunity for the discovery of Venus-like worlds at various stages in their evolutionary history. 
For Proxima Centauri b \cite{meadows2018habitability} demonstrates the capabilities of a HabEx 6.5m space telescope with coronagraph that could be similar to the capabilities of IROV. The inner working angle (IWA) is wavelength dependent and for the HabEx 6.5m they calculate the optimal IWA=1$\lambda$/D=1.17$\mu$m, but in fact the diffraction limit should be 1.22 instead of 1 and this gives 0.96$\mu$m. 
Examining the estimated reflection spectra in  Figures 21--26 in \cite{meadows2018habitability} it is apparent that this instrument may be able to distinguish between 10 bar O2 rich atmospheres, a 90 bar cloud covered Venus, Archean and modern Earth. Both \cite{meadows2018habitability} and \cite{Turbet2016} provide simulations for Proxima Centauri b as both temperate and Venus-like. \cite{Barnes2018}  also demonstrated that it is possible for Proxima Centauri b to have a Venus-like evolutionary path, so our closest neighbor may be denuded, an exo-Earth or even an exo-Venus.

Finally, there is currently a mission proposal to ESA called LIFE \citep{Konrad2021}\footnote{https://www.life-space-mission.com}, which would entail a space based nulling interferometer. This is more-or-less a scaled down and more affordable version of one of the Terrestrial Planet Finder concept missions from nearly two decades ago \citep[e.g.][]{Coulter2003}.

As mentioned above, only one next generation large ($>$30m) optical ground based telescope is fully funded today, so we focus the rest of this section on what  \change{that telescope}{the ELT} will deliver for exoplanetary investigations with applications to exo-Venuses.

There are presently seven different first generation instruments intended for use with the ELT\footnote{https://elt.eso.org/instrument}. Below we focus on three of the first generation instruments relevant to exo-Venus observations (see Table \ref{tab:ELT}). 

HARMONI (High Angular Resolution Monolithic Optical and Near-infrared Integral field spectrograph) \citep{Rodrigues2018,Houlle2021} and METIS (Mid-infrared ELT Imager and Spectrograph) \citep{Brandl2018} are funded via the telescope construction budget while HIRES (HIgh REsolution Spectrograph) \citep{Marconi2018,Marconi2021} is funded by a consortium. We note that HIRES has been renamed ANDES (ArmazoNes high Dispersion Echelle Spectrograph)\footnote{https://elt.eso.org/instrument/ANDES/}, but the instrument architecture remains the same (we will use both names herein).

METIS will operate at 3--19$\mu$m and will focus on high contrast imaging/spectroscopy, along with high spectral resolution integral field unit (IFU) observations. METIS is designed with a coronagraph which will reduce the brightness of an axially-symmetric source (star) by $\sim$ 10$^{-5}$--10$^{-7}$. Low resolution spectra will be obtained with the remaining reflected light for attempted characterization of planets more than 3 Astronomical Units in distance. METIS' IFU mode will have a \SI{1.0}{\arcsecond}$\times$\SI{0.5}{\arcsecond} field of view and will allow for 3km s$^{-1}$ spectral resolution over 2.9--5.3$\mu$m with an angular resolution down to \SI{0.02}{\arcsecond}. METIS will also be capable of direct imaging in thermal emission which will be useful for detecting targets around Sun-like stars where the contrast is less than that of M-dwarfs (mid-IR is 10$^{-7}$ while 10$^-10$ in the visible) although the yield estimates are at most a few such objects \citep{Quanz2015,Bowens2021}. 

The near infrared arm of the HIRES instrument is a more capable version of the present day European Southern Observatory (ESO) Very Large Telescope (VLT) CRIRES+ (The CRyogenic InfraRed Echelle Spectrograph Upgrade Project) instrument\footnote{https://www.eso.org/sci/facilities/develop/instruments/crires\_up.html} for transmission spectroscopy. Baseline wavelength coverage is expected to be 0.55—1.80 $\mu$m with a goal of 0.33—2.44 $\mu$m at a spectral resolution 100000--150000, the bigger mirror allowing higher resolution studies than with CRIRES+. With the Integral Field Unit (IFU) HIRES will observe reflection spectra of nearby exo-Venus candidates discovered via transits, and radial velocity (RV) surveys. Given the geometrical constraints of transiting candidates many more nearby candidates will be available via RV surveys. Figure 2 of \cite{Lovis2022} depicts the possible reflected light candidates for two different IWAs for ELT at 0.75 and 1.5$\mu$m. Although the TRAPPIST-1 planets \citep{Gillon2016} are beyond the reach of HIRES reflection spectroscopy because they are within the IWA, they will be accessible via transmission spectroscopy. 

Given their capabilities for transmission, thermal and reflection spectra HIRES and METIS should allow us to disentangle the atmospheric chemical composition of exo-Venuses and exo-Earths within the habitable and possibly Venus zones (e.g. as shown for the Proxima Centauri b system by \citealt{Turbet2016,meadows2018habitability}) for nearby exoplanetary systems. They may be capable of catching a young exo-Venus in its magma ocean/steam atmosphere phase \citep[e.g.][]{Martins2013,Kawahara2014}, possibly helping to constrain modelling studies \citep[e.g.][]{Matsui1986,Elkins-Tanton2008,Hamano2013,Lebrun2013,Salvador2017,Turbet2021}.

HARMONI will leverage a combination of adaptive optics, a high-contrast imaging module, a medium resolution IFU (R up to 17 000) and a coronagraph to study exoplanets. The approach was first described by \cite{Sparks_Ford2002} and in 2015 \cite{Snellen2015} demonstrated the potential for this combination for the ELT. \cite{Hoeijmakers2018} used a medium resolution IFS on the VLT SINFONI instrument \citep{Eisenhauer2003} similar in many respects to HARMONI (but without a coronagraph) to characterize $\beta$ Pic b. Hence the HARMONI instrument coupled to the ELT has tremendous potential for exo-Venus characterization. It is worth mentioning that a second generation high-contrast imager called PCS has been proposed for the ELT \citep{Kasper2021}. PCS would combine extreme adaptive optics with high spectral resolution exploiting the full potential of this technique on the ELT.

\begin{table}[ht!]
\scriptsize
\caption{First generation ELT instruments relevant to exo-Venus characterization.}
\label{tab:ELT}
\begin{tabular}{|l|p{1.2in}|p{0.7in}|c|p{1in}|}
\hline
Instrument & \multicolumn{3}{|c|}{Main Specifications} & Exo-Venus Science\\
\hline
           & Field of view & Spectral   & Wavelength& \\
           & slit length   & resolution & coverage  & \\
           & pixel scale   &            & ($\mu$m)  & \\
\hline
           & Imager+coronagraph&          &      & \\
           & $\sim$10x10"       & L,M,N +&&\\
           & @ 5 mas/pix in L,M & narrowbands& 3--19  & Thermal Emission \\
           & @ 7 mas/pix in N    &  & &  \\
           &&&&\\
METIS      & Single slit & R$\sim$ 1400 in L, 1900 in M, 400 in N. & 3--13 & \\
           &&&&\\
           & IFU 0.6x0.9"  &&&\\
           & @ 8 mas/pix & L, M Bands && Transmission \& \\ 
           & w/coronagraph            &  R $\sim$ 100 000 & 2.9--5.3 & Reflection Spectra\\
           \hline
           & IFU 4 spaxel scales     &R$\sim$3200 & & \\
HARMONI &0.8x0.6" @ 4mas/pix      &R$\sim$7100 & 0.47--2.45 & Reflection Spectra \\
         &  6x9" @ 30x60mas/pix  &R$\sim$17 000 & & \\
          & (w/coronagraph)           &              & & \\
\hline
ANDES/  & Single Object & R$\sim$100 000 & 0.4-1.8 & Transmission \& \\
HIRES   & IFU (SCAO)    & R$\sim$100 000 &  " & Reflection Spectra\\
\hline
\end{tabular}
\end{table}

It may be possible to image accreting exoplanets in IR wavelengths \citep{Mamajek2007,Miller-Ricci2009,Bonati2019}.
\cite{Miller-Ricci2009} predicted several near infrared windows that would allow detection of a magma ocean. 
However, if water vapor is a major component of the atmosphere (which is not a given, see work by e.g. \citealt{Bower2022}) \citet[][see Supplementary Information]{Goldblatt2013} has shown that the atmosphere may be opaque at most optical and IR wavelengths making characterization problematic.
As mentioned above, the ELT HIRES \& METIS instruments may have the capabilities to characterize not only the magma ocean and steam atmospheres  \citep[e.g.][]{Lupu2014,Hamano2015,Bonati2019}, but may also tell us if modelling studies of a temperate Venus \citep{Way2016,Way2020} are correct to place it in the habitable zone in its early history. The study by \cite{Bonati2019} points to a K-band window around 2.2$\mu$m being optimal at ELT with the smallest inner working angle of 24 milliarcseconds, but calculations by \cite{Turbet2021} could imply that the shorter wavelengths offered by HIRES may prove sufficient.

A number of studies have shown that it may be possible to detect the rotation rate, and other surface features such as ocean glint from single pixel images or low resolution spectroscopy of exoplanets
\citep[e.g.][]{Palle2008,Robinson2014,Fujii2014,Lustig-Yaeger2018,Jiang2018,Gomez-Leal2016,Mettler2020,Ryan2021,Li2021}. Rotation rate in particular has direct application to Venusian studies. Venus' present day retrograde rotation rate and how it might have come about has been studied for decades \citep[see][for a review]{vanHoolst2015}. A variety of explanations have been put forward for its present-day obliquity and slow rotation rate, from impactors \citep[e.g.][]{McCord1968}, solid-body tidal dissipation \citep[e.g.][]{MacDonald1964,Goldreich1966,Way2020}, core-mantle friction \citep{Goldreich_Peale1970,Correia_Laskar2001,Correia2003a,Correia2003b}, oceanic tidal dissipation \citep{Green2019}, to atmospheric tides \citep{Ingersoll1978,Dobrovolskis1980a,Dobrovolskis1980b,Dobrovolskis1983}. Investigators have used Earth observation satellites, such as DSCOVR\footnote{https://solarsystem.nasa.gov/missions/DSCOVR}\citep{Jiang2018}, and space missions such as EPOXI\footnote{https://www.nasa.gov/mission\_pages/epoxi} \citep{Robinson2014} for exoplanetary purposes. For example, DSCOVR has a charged coupled device array 2048$\times$2048 pixels with sizes of 15 $\mu$m. Wavelength coverage is from 200 to 950 nanometers. \cite{Jiang2018} shrank the DSCOVR high-resolution 2-D images down to a single pixel and successfully extracted estimates of the land/ocean ratio and rotation rate. This implies that with a sufficient cadence, the same single pixel `images' we obtain for exoplanets may allow us to constrain their rotation rate \citep{Li2021} and possibly land/sea ratio. \cite{Robinson2010} also demonstrated that it may be possible to use JWST to detect ocean glint in single pixel images of extrasolar planets\add{, but would require an external occulter which is not available}. With similar techniques, we can hope to get better statistical constraints on exo-Venus rotation rates. We could also gain new insight on the causes behind Venus' present-day rotation rate and what it might have been in the distant past. The importance of discerning the rotation rate of planets in the VZ cannot be understated as it can be tied back to the slowly rotating cloud-albedo feedback seen in GCM models that may allow temperate climates under high insolations as discussed in Section \ref{sec:01b1}. As well, observing glint in an planet in the VZ would also be an important discovery as it would show that VZ planets do exist in the liquid water habitable zone \citep[e.g.]{Kasting1993b,Kopparapu2013}. On the other hand if no glint nor cloud-albedo feedback is seen in slow rotators in the VZ then this would make a good case for Venus never having been in the habitable zone.

\subsection{The importance of primordial \& basal magma oceans}\label{sec:VenusMagmaOcean}

Magma oceans are likely ubiquitous during the early history of terrestrial planets. During the accretion of Venus-sized planets, the gravitational energy released from gathering their mass is sufficient to melt their entire mantles \citep[e.g.][and references therein]{Elkins-Tanton2012}.
Giant impacts can provide \change{even more}{additional} energy. Early mantle melting is also favored by radiogenic heating of short-lived isotopes \citep{Merk2002}, the loss of potential energy during core formation \citep{Sasaki1986, Samuel.etal2010} and by tidal heating if one or several moons orbit the planet \citep{Zahnle2007}. Additional energy sources are available for planets that orbit close to their parent stars (e.g., in the Venus Zone around M dwarfs), including star-planet tidal heating \citep[e.g.][]{Driscoll2015} and, speculatively, magnetic induction \citep[e.g.][]{Kislyakova2017}. Observations of young exoplanets can help test several hypotheses about the early atmosphere and magma ocean of Venus-like planets.

\citet[][this issue]{Salvador2022,Gillmann2022} contain a detailed discussion on Venus' primordial and basal magma oceans. Briefly stated, historical models assumed that Earth and Venus had primordial magma oceans that were overlain by an outgassed, dense atmosphere mostly consisting of H$_2$O and CO$_2$ \citep{Arrhenius1974,Jakosky1979}. As reviewed in \cite{Massol2016}, the idea of a steam \& CO$_2$ magma ocean atmosphere continued to be the dominant hypothesis, although recent work has begun to question the simplicity of this formulation \citep{Lichtenberg2021,Bower2022,Gaillard2022redox}. Several 1-D models provide predictions about the longevity of the magma ocean in relation to the distance of Venus from its host-star \citep{Matsui1986,Elkins-Tanton2008,Hamano2013,Lebrun2013,Salvador2017}, but cannot conclusively constrain the timescale of the blanketing atmosphere. Either Venus' magma ocean was short-lived like that of Earth ($\sim$ 1 Myr), allowing water to condense on the surface, or so long ($\sim$ 100Myr) that the steam atmosphere is photodissociated, with hydrogen loss via atmospheric escape and oxygen absorption by the magma ocean \citep[see][this issue]{Westall2022,Salvador2022}. Recent 3-D atmospheric modelling by \cite{Turbet2021} has shown that the steam atmosphere and subsequent magma ocean lifetime could be long, leading again to a desiccated atmosphere during the magma ocean phase. Their model examined N$_2$, H$_2$O and CO$_2$ constituents from 1--30 bar in partial pressure. While these results should be confirmed by another 3-D GCM, their importance cannot be overstated, as it may determine whether Venus kept most of its primordial water or not, and whether water ever condensed on the surface of Venus. See \citet[][this issue]{Salvador2022} for a more detailed discussion.\newline

To inform studies of Venus, scientists should seek to determine how atmospheric properties vary with the intensity of incident starlight, especially for very young exoplanets. If models that feature an early steam atmosphere for Venus are correct, then we should expect to find steam atmospheres around Venus-like exoplanets that are $<$100 Myr old \citep[see][this issue]{Salvador2022}.
Under some critical threshold of stellar insolation, steam atmospheres may quickly condense into surface oceans. For example, \cite{Turbet2021} suggested that this threshold was ~92\% of Earth’s present-day insolation, meaning that Earth narrowly escaped a Venusian fate. However, this critical value can vary depending on the details of the atmospheric model and uncertain parameters \citep{Hamano2013,Lebrun2013,Goldblatt2013,Kopparapu2013}. The predicted mass and composition of the magma ocean atmosphere results from the partitioning of volatile elements between the melt and the gas phase which is primarily controlled by their solubility within the melt and depends on the redox state of the magma ocean and thus the bulk composition of the exoplanet \cite[e.g.][]{Katyal2020,Barth2020}. Observations of stellar composition can provide meaningful, but not exact, constraints on the compositions of terrestrial exoplanets \cite[e.g.][]{Hinkel2018,adibekyan2021}. While magma ocean outgassing is generally thought to be efficient because of the vigorous convection and associated velocities, other mechanisms, such as interstitial trapping of volatile-rich melt \citep{Hier-Majumder2017}, could drastically alter this view and result in alternative outgassing scenarios \citep[e.g.,][]{Ikoma2018}. Furthermore, the convective dynamics and associated patterns might significantly increase the degassing timescales \citep{Salvador2022_MO}. Then, magma ocean \add{degassing} efficiency would decrease with the planet size and increase with the initial water content. Because of its thermal blanketing effect, the outgassing rate of the atmosphere might strongly affect the cooling of the magma ocean and lead to divergent planetary evolution paths and resulting surface conditions. Many other parameters affecting mantle evolution and mixing such as the rotation rate or the crystallization sequence could significantly affect the volatile distribution and resulting outgassing with time. Yet, they have been poorly studied in the frame of volatile degassing. Thus a complete understanding of the interplay between magma ocean cooling rate, outgassing and their influence on post-MO mantle convection regime and surface conditions is still lacking. Ultimately, a large sample size of exoplanets is needed to derive statistical conclusions.

Detailed characterization of terrestrial exoplanets will remain difficult for at least the next decade. \cite{Schaefer2021} provide a summary of some technical pitfalls. However, some hot, bright planets that orbit very close to their parent stars can be studied with modern technology. For example, observations of the infrared phase curve of the terrestrial exoplanet LHS 3844b, collected with the Spitzer Space Telescope, revealed that it does not have a substantial atmosphere \citep[e.g.][]{Kreidberg2019}, which is consistent with a volatile-poor bulk composition \cite[e.g.][]{Kane2020} or with low outgassing rates. Future observatories could potentially use the direct imaging technique to detect superficial magma oceans for planets that also have thin or nonexistent atmospheres \citep{Bonati2019}. Alternatively, planets with huge amounts of outgassing from a magma ocean might have an atmosphere that is thick enough to affect mass-radius measurements \citep{Bower2019}. In the same way, the partition of water between the atmosphere and the magma ocean of water-rich exoplanets can affects their calculated radii by up to 16$\%$ in some cases \citep{dorn2021hidden}, which would be enough to be tested for close-in bodies, and help understand the evolution of water budget in terrestrial planets. Furthermore, planets sustaining relatively long ($\sim$100 Myr) magma ocean states under a runaway greenhouse due to their proximity to the host star \citep[][type-II planets]{Hamano2013} might also be distinguishable by a radius inflation effect \citep{Turbet2019, Turbet2020}, thus providing additional constraints. In the history of exoplanetary studies, planets with extreme properties (e.g., hot Jupiters) were often the easiest and thus the earliest to be studied. Significant technical advances are needed to explore true exoplanetary analogues to Earth and Venus (see Section \ref{sec:Future_Space_Ground}).

\section{How can Venus inform exoplanetary studies}\label{sec:01b.6}

Our nearest planetary neighbor provides one of the end members of terrestrial habitability in our solar system. With its thick present-day atmosphere and inhospitable surface conditions, Venus is considered to be \change{to}{too} close to our sun to be within the habitable zone, but was Venus ever within the habitable zone? The latter concept would be surprising to any modern-day climate scientist. How can a world that was receiving, 4.2 billion years ago, 1.4 times the incident solar radiation that Earth receives today be inside the habitable zone?  As discussed above and in \cite[e.g.][this issue]{Westall2022}, an efficient cloud albedo feedback from a slowly rotating Venus may have kept ancient Venus temperate according to GCM modeling \citep{Yang2014,Way2016} assuming sufficient surface liquid water and a short lived magma ocean phase \citep{Hamano2013}. If these GCM results are correct, we can expect to find habitable worlds well within the VZ around G-dwarf stars. For planets in the VZ of M-dwarfs, GCM results demonstrate severe limitations in the greater than modern-day Earth solar insolations (1361 W m$^{-2}$) allowed by the redder spectral energy distribution of such host stars \citep{Kane2018}. This is because Earth-like atmospheres are highly efficient at absorbing and trapping the infrared radiation of M-dwarfs, preventing the high insolations and temperate climates seen in GCM exoplanet modelling studies of VZ planets around G-dwarfs \citep{Yang2014,Way2018}. 
As well, the (likely tidally-locked) planets around low mass stars tend to ``rotate" much faster (i.e. shorter orbital periods) than around more massive stars. This results in a reduced cloud albedo feedback at the substellar point \citep[e.g.][]{Kopparapu2017}.
Venus can also become a point of reference when it comes to the behaviour of its interior. For example, it is still debated if Venus' mantle convection is indeed in a stagnant lid regime at present-day, as has long been theorized \citep{Solomatov2004}. However, Venus provides many more clues about the state of its mantle than any exoplanet, and can help discriminate between the multiple scenarios highlighted by numerical studies \citep{ballmer2021diversity}. Finally, most mechanisms at work on Venus (or Earth), are likely to also affect exoplanets, in one form or another.
Venus' ability to inform exoplanetary studies goes beyond providing us with an example of the atmospheric signature of a planet in a runaway greenhouse state with an inhospitable climate: Venus can also help us understand planetary evolution more generally.
For these reasons it is important to understand how our present-day and near-future understanding of Venus can inform the study of exo-Venuses. In the rest of this chapter, we will provide an overview of our understanding of Venus through time.

\subsection{Volatile cycling and weathering on Venus through time}\label{sec:VenusVolatiles}

In addition to a thick, CO$_2$-dominated atmosphere, resulting in an extremely hot climate, Venus also lacks modern Earth-style plate tectonics \citep[e.g.][]{Breuer_Moore2007} and a strong, intrinsic magnetic field. The exact style of tectonics Venus currently exhibits is not well known, due, in large part, to the difficulty in mapping the Venusian surface in sufficient detail. Venus does not appear to fall neatly within either the plate-tectonic or stagnant-lid end-member regimes of tectonics. Although there is no evidence for a global network of plate boundaries and mobile plates, there are regions of the Venusian surface with features strikingly similar to subduction zones on Earth \citep[e.g.][]{Davaille2017,Gerya2014_plume,Sandwell1992}. Moreover, there is evidence for the motion of discrete crustal blocks on Venus, though it is difficult to constrain when this motion may have occurred during Venusian history \citep{Byrne2021}. Finally, Venus' lithosphere is estimated to be thinner than what would be expected if the planet were in a stagnant-lid state \citep{Borrelli2021}.

These significant differences in the magnetospheric, tectonic, and climatic state of Venus compared to Earth also possibly led to significant differences in atmospheric retention, surface weathering, and volatile cycling. Understanding these differences is crucial for interpreting future atmospheric observations from exoplanets, in particular those in the ``Venus zone" \citep{Kane2014} that are thus likely to also be in a runaway greenhouse state. In this section, we will explore how Venus' current state leads to different weathering, volatile cycling, and atmospheric retention processes and behavior than operate on Earth.

Like all rocky planets, Venus' climate is likely coupled to the interior \citep[e.g.][]{gillmann2014atmosphere} and the magnetosphere \citep[e.g.][]{Foley2016}. The hot, thick CO$_2$ greenhouse climate may be both a cause and a consequence of Venus' lack of plate tectonics. Likewise, the presence or absence of a magnetic field may be controlled by the style of tectonics the planet exhibits. Meanwhile, atmospheric evolution is influenced by the magnetosphere, which alters rates of atmospheric escape (See Section \ref{sec:VenusMagFieldEscape}). Such atmospheric evolution then affects the climate, feeding back to interior processes (see Chapter 3b for more). 

Coupling between surface and interior opens up further questions about the evolution of Venus and how it informs exoplanet studies. Do planets that experience a runaway greenhouse necessarily also lose plate tectonics and the operation of a core dynamo? Are runaway greenhouse climates, and their subsequent impact on a planet's interior always externally driven (e.g. due to changes in stellar luminosity), or can they be internally driven as well (e.g. due to changes in tectonics or rates of volatile outgassing via volcanism)? Are the current surface conditions inherited from the cooling of an early magma ocean stage or the results of the long-term evolution? Studying Venus' history can help shed light on these questions. We therefore structure this section as follows: first, we outline the weathering, and volatile cycling that operate on Venus today; next, we discuss how these processes might have evolved throughout Venusian history, and what constraints we have on this evolution; finally, we discuss how these processes are coupled to the interior evolution, and how this coupling could dictate rocky planet evolution in general. 

\subsubsection{Volatile cycling and weathering on present-day Venus}\label{sec:volatile_cycling_weathering}

Volatile cycling on Earth is driven by volcanic outgassing from the interior and weathering processes, which reincorporate outgassed volatiles into rocks at the surface. The latter is typically facilitated by water-rock reactions, and ingassing of volatiles via the return of these volatilized surface rocks to the interior, typically through subduction. On Venus, the extremely hot climate, lack of liquid water at the surface, and lack of global-scale plate tectonics means volatile cycling, to the extent it can occur, must behave very differently than on Earth. 

Some of the key volatiles for the evolution of Venus' atmosphere and surface environment are C, H, N, and S. Considering C \& H first, there is a clear dichotomy in these species at the surface and in the atmosphere between Earth \& Venus today: Venus' surface is dry and the atmosphere is dominated by $\sim$90 bars of CO$_2$ \citep[e.g.][]{Mogul2023}, while on Earth liquid water is abundant and CO$_2$ is only a trace gas in the atmosphere. This dichotomy leads to significant differences in weathering, but may also have been caused by differences in weathering. 

\subsubsection{Weathering}\label{Weathering}

On Earth, the carbonate-silicate cycle operates to regulate the amount of CO$_2$ in the atmosphere, and maintain a temperate climate throughout most of Earth's history \citep[e.g.][]{Walker1981,Berner1993,Kasting1993a}. Silicate weathering is the primary mechanism for removing CO$_2$ from the atmosphere in this cycle, and the dependence of the rate of silicate weathering on climate state creates a negative feedback. Weathering on the modern Earth is driven by reactions between exposed rock on Earth's surface, as well as rock on the seafloor near mid-ocean ridges \citep[e.g.][]{Brady1997,Coogan2013,Coogan2015,KrissansenTotton2018}, and CO$_2$ dissolved in rainwater and the oceans. Liquid water is therefore critical, and weathering will be severely limited on a planet lacking liquid water, like Venus. There is some chemical reaction between Venus' CO$_2$-rich atmosphere and surface rocks (See Chapter 3b for a detailed discussion), as evidenced by carbonate-rich coatings, which may form as an intermediate step in weathering of Venus' surface \citep{Dyar2021}.  Nevertheless, the slow gas-solid reactions and the limited erosion in the absence of water prevents the efficient consumption of atmospheric CO$_2$ by the formation of carbonates \citep{Zolotov2019}. In addition, carbonates are thermodynamically unstable at Venus' surface, where they react with sulfur species, in particular SO$_2$, from the atmosphere to form sulfates \citep{Gilmore2017}. Indeed, the elevated bulk sulfur content of 0.65±0.40wt\% and 1.9±0.6wt\% recorded at the Venera 13 and Vega 2 landing sites, respectively \citep{Surkov1984,Surkov1986} indicates net trapping of sulfur-bearing phases from the atmosphere into surface rocks \citep{Zolotov2019}. All told, the lack of liquid water on Venus today means that weathering cannot act as an efficient removal process for atmospheric CO$_2$.

Such inefficient silicate weathering could in fact partly explain why Venus' present-day atmosphere is CO$_2$ dominated. Without weathering to remove it, CO$_2$ continuously accumulates in the atmosphere, as volcanic degassing from the interior proceeds. Earth contains a similar amount of CO$_2$ locked in carbonate rocks as exists in the Venusian atmosphere today \citep[e.g.][]{Ronov1969, Holland1978, Lecuyer2000}, thanks to active weathering processes on the Earth. 

Another key factor is that weathering on Earth is also tied to tectonics. For weathering to be continuously active, erosion is needed to transport weathered rock away, and expose fresh rock. In the extreme case where there is no erosion whatsoever, weathering would cease entirely once a layer of weathered rock formed at the surface, as ground water would be unable to reach fresh, weatherable rock. A less extreme, and more common scenario, is when the rate of silicate weathering becomes limited by the supply of fresh rock brought to the near surface environment by erosion. In this case, all climate feedback involved in silicate weathering is lost; the weathering rate depends only on the erosion rate, as all fresh rock is weathered nearly instantly when brought into the weathering zone near the surface. Weathering reaching this state of being globally ``supply limited" is another potential mechanism for forming a CO$_2$ dominated, hothouse climate, even if liquid water is still present on a planet's surface \citep[e.g.][]{Foley2015,Kump2018}. 

Silicate weathering is also linked to the land area of the planet: Wind and rainfall on emerged continents promote erosion and, in turn, the rate at which new surface is exposed. A large land area is however not vital for a stable climate: On a planet largely covered by oceans, seafloor-weathering dominates and can regulate the atmospheric CO$_2$ to some extent \cite[e.g.][]{Foley2015,Hoening2019,KrissansenTotton2018}.

As erosion rates are ultimately bounded by rates of tectonic uplift, it has been previously argued that plate tectonics might be essential for silicate weathering \citep[e.g.][]{Kasting2003}. As a result, another possible explanation for Venus' present-day atmospheric state could be that a lack of plate tectonics limits silicate weathering, allowing volcanically outgassed CO$_2$ to build up in the atmosphere. However, even without plate tectonics there are processes, such as volcanism, that act to supply weatherable rock to the surface. So whether a lack of plate tectonics leads to a hothouse climate depends on whether these other processes can supply enough fresh, weatherable rock to keep pace with CO$_2$ outgassing. \cite{Foley2018} argue that even in a stagnant-lid regime, volcanism provides a sufficient supply of weatherable rock to sustain temperate climates. This study considered outgassing of CO$_2$ from the mantle and from decarbonation of crustal carbonate as it is buried by fresh lava flows, and found that a much higher concentration of CO$_2$ in erupted magma than on the modern Earth would be needed for a hothouse climate to form.
However, the amount of CO$_2$ outgassed also depends on the types of materials through which magmas penetrate on their way to eruption \citep[e.g.][]{Henehan2016}. If magmas erupt through C-rich crustal rocks, more CO$_2$ can be released than one would expect based on mantle CO$_2$ concentration alone. For example, in the case of the Siberian Traps, volatile release likely outweighed weathering as a result of magma interaction with crustal rocks \citep[e.g.][]{Svensen2009}. However, such high CO$_2$ degassing rates may be anomalous and, geologically speaking, short-lived, as they require magmas to first hit regions where crustal rocks are C-rich, and then can only be maintained until these pockets of C-rich crustal rocks have been exhausted. Maintaining a permanent hothouse climate with liquid water present would require CO$_2$ degassing rates to continuously exceed silicate weathering rates through the planet's lifetime. 

It therefore remains unclear exactly how the present atmosphere of Venus came about  if there was an earlier temperate period \citep{Head2021}. A loss of water due to a runaway greenhouse climate would almost certainly lead to the buildup of a thick CO$_2$ atmosphere, as long as volcanism was still active. A lack of plate tectonics, with liquid water still present, could impede weathering to the point where a hothouse climate forms, but this would require either a CO$_2$-rich mantle or for magmas to interact with C-rich rocks as they erupt; without either of these two conditions weathering can still maintain a temperate climate even in a stagnant-lid regime of tectonics. 

Whether the tectonic regime or the presence of liquid water is the more significant limitation on weathering processes has important implications for exoplanets. If weathering is not strongly affected by tectonic regime, then one does not need to know a planet's tectonic regime in order to assess whether a carbonate-silicate cycle, capable of sustaining habitable surface conditions, can operate. Estimating an exoplanet's tectonic state from remote observations will be a significant challenge, so testing whether habitability is possible without plate tectonics is critical for exoplanet studies. 
Future Venusian exploration can help test the importance of tectonics for weathering and habitability. If Venus is shown to have had active silicate weathering in the past, while also lacking plate tectonics, then we would have direct evidence that plate tectonics is not necessary for the carbonate-silicate cycle. On the other hand, if Venus' history indicates the loss of water through a runaway greenhouse was the primary causal factor for Venus' CO$_2$-rich atmosphere, then we'd expect exoplanets that have experienced runaway greenhouses to have similar atmospheric states. Such expectations can be tested with future observations, as outlined in Section 1. Going further, exploring when and why the carbonate-silicate cycle ultimately failed to regulate the climate on Venus, as must have happened at some point during Venus' history, would offer clues to the conditions for habitability of terrestrial planets \citep[see also][this issue]{Westall2022}.

\subsubsection{Volcanism \& Outgassing}\label{Volcanism_and_Outgassing}

Weathering is not the only aspect of the carbonate-silicate cycle that is essential for regulating atmospheric CO$_2$ levels. Volcanic outgassing is also necessary, at sufficiently high rates, to maintain enough CO$_2$ to prevent global glaciation \citep[e.g.][]{Walker1981,Kadoya2014,Foley2018,Stewart2019}. 
Venus today is of course near the other extreme limit, with a CO$_2$ dominated atmosphere, rather than a CO$_2$ poor one. However, the importance of volcanic outgassing to rocky planets in general highlights the question of whether Venus is actively outgassing today. 

The variations of SO$_2$ in the atmosphere of Venus have been recorded by Venera 12 \citep{Gelman1979}, Pioneer Venus \citep{Oyama1980,Esposito1984} and Venus Express \citep{Marcq2013}. Combined with models these can give estimates of the column sulfur abundance \citep[e.g.][]{Schulze2004,Krasnopolsky2016}. The variations of SO$_2$ and the maintenance of the H$_2$SO$_4$ cloud layer on Venus have been suggested to indicate volcanic activity. Since SO$_2$ reacts with calcite (CaCO$_3$) on the surface of Venus to form anhydrite (CaSO$_4$), it will be consumed unless replenished by volcanism. 
Following \cite{Gilmore2017} this can be written as CaCO$_3$(calcite)+1.5 SO$_2$(gas)$\rightarrow$CaSO$_4$(anhydrite)+CO$_2$(gas)+0.25 S$_2$(gas).
\citet{Fegley1989} 
calculated a sulphur removal rate of $2.8\times10^{13}$ g yr$^{-1}$. In order to  maintain the global H$_2$SO$_4$ cloud layer, this removal rate needs to be balanced by a volcanic outgassing rate of $5.6\times10^{13}$ g yr$^{-1}$ or 1.1 Pa kyr$^{-1}$ SO$_2$. Depending on the S/Si ratio of erupted material, \citet{Fegley1989} estimated the equivalent global volcanic eruption rate to 0.4 -- 11 km$^3$/yr.
This rate is lower than the total average output rates on Earth of about 26 -- 34 km$^3$/yr, of which about 75\% are contributed by ocean-ridge magmatism \citep{Crisp1984}, while recent work by \cite{Byrne2022} implies that Venusian volcanic rates should be similar to those on modern Earth.
It should be noted, however, that atmospheric dynamics and chemistry may be responsible for the variability of sulfur species in the atmosphere of Venus \citep{Hashimoto2005, Marcq2013}. The measurements mentioned above will be improved upon with mass spectrometer observations from the upcoming DAVINCI mission \citep{Garvin2020}\footnote{https://www.nasa.gov/feature/goddard/2021/nasa-to-explore-divergent-fate-of-earth-s-mysterious-twin-with-goddard-s-davinci} which will help to better constrain column abundances of sulphur and a number of other species. As well, the DAVINCI in-situ infrared (IR) imaging camera should help connect surface observables to the orbiting IR and radar instruments on VERITAS and Envision \citep{Widemann2022}
to confirm or refute previous indications of on-going volcanism \citep[e.g.][]{Smrekar2010,Shalygin2015,Gilmore2017} as a possible sulfur source, and provide valuable insight to exoplanet studies. 

Remote observations of H$_2$O and HDO have been made from Venus orbit \citep[e.g.][]{Cottini2012}, from Earth ground based instruments  \citep[e.g.][]{Encrenaz1995,Sandor2005}, and from in-situ instruments on the Pioneer Venus large probe and Venera 15 \citep{Donahue1982,Koukouli2005}. A compilation of H$_2$O measurements by \cite{DeBergh2006} gives atmospheric column values from 20--45ppmv with one measurement at 200ppmv. It is generally assumed that H$_2$O sources are volcanic like those of its sulphur counterparts \citep[e.g.][]{Fegley2003,Fegley2014,Truong2021}. 

Tying the abundances of N$_2$ in the upper atmosphere to lower atmosphere abundances remains challenging \citep[e.g.][]{Peplowski2020}. N$_2$ as the second most abundant gas in the Venusian atmosphere is often overlooked, but it corresponds to nearly four times the atmospheric abundance on Earth when scaled by planetary mass. Here again the DAVINCI mission will give more accurate column abundances of N$_2$ and in combination with photochemical modelling \citep[e.g.][]{Krasnopolsky2012} may help us to better understand the upper atmosphere abundances and how those tie to possible surface sources and the N$_2$ cycle in general. N$_2$ is certainly a challenging gas to detect in exoplanetary atmospheres, but \cite{Schwieterman2015} has shown that it may be possible.

Future atmospheric characterization of exoplanets can also help test models of volcanic outgassing, by potentially identifying 
ongoing volcanic activity on such planets. SO$_2$ has been proposed as a proxy for explosive volcanism \citep{Kaltenegger2010}, as well as sulfate aerosols \citep{Misra2015}. Sulfate aerosols are formed during volcanic eruptions and have a lifetime of months to years in the atmosphere; as such they may be detectable in transit transmission spectra \citep{Misra2015}. Venusian measurements are critical to helping us constrain the longevity and rate of volcanism on rocky exoplanets -- a key question for interpreting future atmospheric observations performed by upcoming missions such as JWST and ELT (See Section \ref{sec:Future_Space_Ground}). Additional modelling studies have investigated volcanism and outgassing of terrestrial exoplanets \citep{Kite2009,Tosi2017,Noack2017,Dorn2018,Foley2018,Foley2019}. These studies provide predictions for how long volcanism can last on planets in different tectonic regimes, with different sizes, heat budgets, and material properties. On Exo-Venus planets with an atmosphere similar to that of Venus, the signal of SO$_2$ and other volcanic gases needs to be detected above an optically thick lower atmosphere. However, volcanic gas plumes are less buoyant in a hot and dense atmosphere and may thus not reach high enough altitudes compared to altitudes reached in otherwise thinner and colder atmospheres \citep{Henning2018}. 

In addition, analogs of present-day Venus may present a featureless spectra both in transit transmission and in direct imaging (See Section \ref{Transmission_Emission_JWST} and Figure \ref{fig:PSG_Venus}), making their characterization difficult \citep{Arney2018,Fauchez2019}. Nevertheless, these challenges further emphasize the necessity of additional Venus exploration. By studying Venus' present-day atmosphere, interaction with any present-day volcanism, and the evolution of the atmosphere over time, we could test these proposed proxies for exoplanetary volcanism, and perhaps develop more effective ones. 

As mentioned above, studying Venus' evolution may help constrain further predictions from models of exoplanet outgassing and climate evolution. For example, in a study employing parameterized thermal evolution modelling and mantle outgassing, \citet{Tosi2017} investigated the habitability of a stagnant lid Earth (an Earth-like planet without plate tectonics) and found that depending on the mantle redox conditions, several hundreds bar of CO$_2$ may be outgassed.
Moreover, models of mantle melting and volatile partitioning suggest that the chemical composition of the atmosphere and the dominant outgassed species are strongly controlled by the redox state of the mantle \citep{Ortenzi2020}.
For sulfur species both fO$_2$ and water content are critical \citep{Gaillard2009, Gaillard2014}. For a given water content, the outgassed sulfur increases for increasing fO$_2$. For oxidising conditions, SO$_2$ is the dominant sulfur species irrespective of the water content. For reduced conditions, however, SO$_2$ and S$_2$ are the dominant sulfur species for hydrated melts \citep{Gaillard2009}.
At the same time surface pressure also affects the final composition of the gases released into the atmosphere. For example, high surface pressures may limit outgassing of water, because the solubility of the latter in surface lava significantly increases for atmospheric pressures larger than 10\,bar \citep{Gaillard2014}.
Under present-day Venus surface pressures, the most dominant outgassing species is CO$_2$, while only a small portion of SO$_2$ and water is expected to be outgassed, due to their high solubility in surface lava \citep{Gaillard2014}. If constraints on Venus' interior oxidation state can be placed by measuring atmospheric H$_2$/H$_2$O and temperature \citep[e.g.][]{Sossi2020}, then results from these models can potentially be tested by both the present-day atmospheric makeup, and whatever constraints on the long-term evolution of the atmospheric composition are developed from future missions. This ability to benchmark outgassing models against Venus will improve our predictions for the atmospheres of exoplanets. Future missions will be used to constrain the present-day atmospheric composition and perhaps surface water abundances. These are particularly interesting as they may be directly related to mantle water abundance which would help constrain the range of water content-dependent parameters associated with mantle melting \citep[e.g.,][]{Hirschmann2006, Ni2016} and convective dynamics such as viscosity and density \citep[e.g.,][]{Lange1994}.

Venus may also be able to help us to predict the evolution and habitability of terrestrial exoplanets more generally. Since most exoplanets detected thus far are larger than Earth and Venus, a scaling of the main physics with planet size and mass is crucial. For Venus-like planets with a similar relative core mass fraction, the planet mass can be directly derived from its size \citep{Valencia2006}. 
When exploring the habitability of massive planets, it is important to attempt to quantify the volcanic outgassing rate which controls the atmospheric partial pressure of CO$_2$ regardless of their tectonic state. On the one hand, the mantle temperature generally increases with the size of a planet, which increases the strength of convection and the melting depth. This favours an increasing outgassing rate with planet size. On the other hand, the pressure gradient is higher in more massive planets, which reduces the strength of convection and the melting depth, favoring smaller outgassing rates of massive planets. The melting depth is particularly important for stagnant-lid planets, since on a planet with plate tectonics, mantle material can rise to the surface at mid-ocean ridges. 
An additional important factor to be considered for massive planets is the buoyancy of partial melt, which needs to be positively buoyant in order to rise to the surface. Since the density of melt increases more strongly with pressure than solid rock, only melt that forms below a certain pressure contributes to volcanic outgassing \citep{Ohtani1995,Agee1998}. The above noted competing mechanisms typically lead to a higher degassing rate for planets between 2 and 4 Earth masses and a reduced outgassing rate for more massive planets \citep{Noack2017,Dorn2018,Kruijver2021}. 
Compared to smaller planets, high outgassing rates of large planets can last longer, since their larger ratio between volume and surface area implies a less efficient cooling. While for massive stagnant-lid planets, the above noted effects can even lead to a cessation of volcanism, \citep{Noack2017,Dorn2018}. This is not the case for planets with plate tectonics where the melting region is extended closer to the surface beneath mid-ocean ridges \citep{Kite2009,Kruijver2021}.

A recent study by \citet{Quick2020} finds that even massive exoplanets such as 55 Cancri e, an 8\,M$_E$ rocky exoplanet, might be volcanically active based on the estimated heat sources (radiogenic and tidal) available in their interior. Rocky exoplanets closely orbiting their parent star may experience volcanic activity focused only on one hemisphere, due to the strong surface temperature variations caused by their tidally locked orbit \citep{Meier2021}. Altogether, understanding physical processes that control volcanic outgassing of Venus throughout its evolution, and studying the sensitivity of these processes to planetary parameters such as size, bulk composition, and tectonic state, will greatly advance our estimates of the atmospheric composition of exoplanets.

\subsubsection{Volatile ingassing}\label{Volatile_ingassing}

As explained in Section \ref{Weathering} silicate weathering can regulate the amount of CO$_2$ in the atmosphere if liquid water is present on the surface. The carbon that is removed from the atmosphere eventually becomes stored in carbonate sediments, which are subsequently buried on the seafloor. The fate of these sediments on longer timescales is controlled by the tectonic regime of the planet. Plate tectonics allow for a relatively shallow temperature-depth gradient in subduction zones, which allows large parts of the carbonates to remain stable during subduction. On modern Earth, approximately half of the carbon that enters subduction zones is released at arc volcanoes, although this fraction strongly depends on the temperature-depth profile of the individual subduction zone \citep{Sleep2001,Dasgupta2010,Ague2014}. The remaining carbon is subducted into the mantle, which closes the deep carbon cycle. On exoplanets with plate tectonics the fraction of subducted carbon that enters the mantle may differ significantly. On planets with higher plate speed, steeper angle of subduction and/or smaller mantle temperature, carbonates would not heat up as strongly during subduction and a larger fraction could remain stable. For example, cooling of the Earth’s mantle during the past 3 Gyr could have enhanced the carbon fraction that enters the mantle by approximately 10\% \citep{Hoening2019}. On timescales of millions to billions of years, this variation can play a key role in the distribution of carbon between the mantle and the atmosphere.

Without plate tectonics, transporting carbon into the mantle is challenging. The slow sinking of carbonated crust, as it becomes buried by new lava flows, results in a thermal equilibrium with the surrounding rock. The bulk of the carbonates becomes unstable at a relatively narrow temperature interval \citep{Foley2018}, which is usually exceeded within the stagnant lid. If the released CO$_2$ is transported with uprising lava or through cracks to the surface, recycling of carbon into the mantle is rare. As a result, the combined crust-atmosphere carbon reservoir on stagnant-lid planets would steadily increase with ongoing volcanic outgassing. Since the release rate of CO$_2$ from the crust into the atmosphere depends on the crustal carbon reservoir, an important consequence is that atmospheric CO$_2$ retains a memory of its initial value.
The initial atmospheric CO$_2$ reservoir may be erased quickly, but if this then gets stored in the crust and is not recycled into the mantle, CO$_2$ release (and therefore atmospheric CO$_2$) in the subsequent evolution would still depend on the initial CO$_2$. However, on planets with plate tectonics, the initial carbon distribution becomes unimportant after some million years \citep{Foley2015}, because of the recycling. 
Another important consequence is that weathering cessation could result in a dramatic rise of atmospheric CO$_2$, since all carbon that has been degassed during the entire history of the planet would accumulate in the atmosphere. In case of early Venus the atmospheric CO$_2$ concentration would have increased by approximately one order of magnitude within 100 Myr \citep{Hoening2021}. 
Altogether, volatile ingassing strongly affects the long-term atmospheric evolution of a planet. Predicting volatile ingassing does not only require knowledge about the tectonic and thermal state of the planet but furthermore a precise understanding of the fate of released CO$_2$ in the crustal matrix.

As explained in Section 2.1, there maybe active subduction in localized regions of Venus today, possibly driven by lithospheric burial under plume-induced volcanism and subsequent rollback of the buried lithosphere \citep{Gerya2014_plume, Davaille2017}. Although the Venusian crust is not highly volatilized today, due to the lack of liquid water and hence nearly non-existent weathering, this style of subduction could potentially drive volatile ingassing if it were active with liquid water present. Rates of ingassing possible with this style of limited subduction have not been well studied, but are likely much lower than ingassing rates that would be seen with Earth-like plate tectonics. Venus exploration can thus potentially help constrain rates of volatile ingassing for planets that lie in between the end-member plate-tectonic and stagnant-lid regimes, and help inform the range of volatile cycling behavior that might be seen on exoplanets.

\cite{bean2017statistical} discussed a comparative planetology approach to test the habitable zone concept: If silicate weathering is generally temperature-dependent on exoplanets with liquid surface water, the atmospheric CO$_2$ concentration on the planet should decrease with increasing incident insolation, for example as a function of stellar type, age, distance between the star and the planet. When incident insolation exceeds a critical value, surface water would evaporate and weathering would cease. Therefore, we would expect to observe an abrupt increase of atmospheric CO$_2$ on planets at the inner edge of the habitable zone \citep{Turbet2019b,graham2020thermodynamic}. For stagnant-lid planets, this abrupt CO$_2$ increase might even be more pronounced, because volcanic degassing would be accompanied by a release of CO$_2$ from buried carbonates. From thermal evolution models coupled to a carbon cycle model for stagnant lid planets, \cite{Hoening2021} predicted an increase of the CO$_2$ concentration on planets at the inner edge of the habitable zone of at least one order of magnitude.

\subsubsection{Weathering and the sulfur cycle on Venus today}\label{Weathering_sulfur_cycle}

The chemical interaction between the surface and atmosphere on Venus is particularly important as it can affect the sulfur cycle \citep[see][this issue]{Gillmann2022}. 
The latter plays a dominant role in the complex photochemistry and dynamics of Venus' atmosphere affecting sulphuric acid cloud formation \citep[e.g.][]{Fegley1989}, the presence of an optically thick aerosol layer \citep{Knollenberg1980} and variations of SO$_2$ atmospheric content \citep{Esposito1984, Marcq2013}. While sulfur and other atmospheric species could be supplied to the atmosphere via volcanic activity, whose present-day level has large uncertainties \citep[][and references therein]{Mueller2017}, weathering processes act as a sink to remove these through complex multiphase chemistry. This is yet another area where exoplanet observations can play an important role in discerning not only the state of the atmosphere in a VZ planet, but may also provide some constraints on volcanic activity for a modern Venus-like world with measureable SO$_2$ abundances.

\subsection{Venus' magnetic field}\label{sec:VenusMagField}

Venus lacks a global (i.e., strong) magnetic field today. As discussed in \citet[][this issue]{ORourke2022}, any intrinsic magnetism in Venus must be relatively weak—specifically producing magnetic fields $\leq$ 5--10 times weaker at the surface than Earth's dynamo-generated field \citep{Phillips_Russell1987}. However, we currently have no meaningful information about the magnetic history of Venus prior to the Mariner 2 flyby in 1962. Understanding why Venus has no global magnetic field now and whether one existed in the past is important for several reasons \citep[e.g.][]{Lapotre2020,Laneuville2020}. 
First, planetary magnetism is intrinsically interesting as a complex phenomenon \citep[e.g.][]{Stevenson2003,Stevenson2010}.
Second, the absence (or presence) of a global magnetic field places constraints on models of planetary formation and thermal evolution. Finally, magnetic fields may play key roles in mediating atmospheric escape processes over time (See Section \ref{sec:VenusMagFieldEscape} below). Studies of Venus provide clues about how magnetic fields will shape the evolution of exoplanets. At the same time, studies of exoplanets may elucidate if the magnetic aspect of the Earth/Venus dichotomy is a natural corollary to the differences in atmospheric conditions--that is, are the prospects for a long-lived, global magnetic field correlated with surface habitability?\newline

Studying planetary magnetism is thus a ``two-way street" between Venus and exoplanets \citep{Lapotre2020}. Over the next few decades, we should advance our scientific understanding by both exploring Venus and searching for extrasolar magnetospheres. Various direct and indirect methods for detecting magnetic fields at exoplanets have been proposed. Space-based radio telescopes could search for direct radio emission \citep[e.g.][]{Driscoll_Olson2011}. Other ideas include searching for various types of auroral emission from exoplanets---or evidence of the interaction of stars and the stellar wind with magnetized exoplanets \citep[e.g.][]{Lazio2016,Vedantham2020,Pope2020}. Brown dwarfs are the current frontier for direct detections of magnetic fields \citep[e.g.][]{Kao2018}. Indirect evidence has been presented for the magnetic fields of hot Jupiters from stellar interactions \citep[e.g.][]{Cauley2019}.

There are a number of geodynamic scenarios for Venus which may have implications for exoplanetary studies. Venus lacks a global magnetic field today because it does not have a strong dynamo operating in its deep interior. Although Venus rotates slowly compared to Earth, a dynamo would still exist if a large amount of electrically conductive liquid were churning vigorously. Such reservoirs (e.g., a metallic core that is at least partially liquid) might exist, but they are currently stagnant. Broadly speaking, two types of scenarios have been proposed to explain why no dynamo operates within Venus. These scenarios make different predictions about whether any crustal remnant magnetism might await detection on Venus. Moreover, these scenarios imply different predictions for what kinds of exoplanets will host global intrinsic magnetic fields.

The first type of story for Venus' magnetic history argues that the tectonic state of Venus prevents any dynamo from operating in the deep interior. As discussed in the previous section (and shown in Fig. \ref{fig:thermalhistory}), the interior of Venus is thought to cool more slowly than Earth's if its mantle operates in the episodic-- and/or stagnant--lid regime. Venus could have a metallic core that has the same bulk composition and is chemically homogeneous, like Earth’s core. However, iron alloys are thermally as well as electrically conductive \citep[e.g.][]{Williams2018}, so thermal conduction can transport all the heat from a slow-cooling core without any fluid motion. Earth’s cooling rate is arguably only somewhat higher than the critical value required to sustain convection \citep[e.g.][]{Nimmo2015,Davies2015,Labrosse2015}. Slow cooling is  thus fatal to the chances for a dynamo in Venus at present-day \citep[e.g.][]{Nimmo2002,DriscollBercovici2014,ORourke2018}. This general conclusion also holds if Venus initially had a basal magma ocean \citep{ORourke2020}. Critically, a dynamo seems more likely to have operated in the past. In this case, crustal remnant magnetism may provide a detectable record of an early dynamo  \citep[e.g.][]{ORourke2019}.  

The second type of story proposes that the stochastic nature of the accretion of Venus doomed the chances for a dynamo from the start. Specifically, \cite{Jacobson2017} proposed that Venus did not suffer a late energetic impact (but see \citealt{Jacobson2022} this issue for the latest research on this topic). The absence of such an impact would mean that the core of Venus could have an onion-like structure where the outermost layers were added last. As proto-Venus grew, its interior grew hotter and had higher pressures. Core-forming material would thus equilibrate with silicates under progressively more extreme conditions, causing more light elements such as silicon and oxygen to partition into the iron alloy \citep[e.g.][]{Siebert2013,Fischer2015}. This process would establish a stable density gradient in the core that prevents convection--material containing a few weight percent of extra light elements would need to cool by thousands of degrees (impossibly) to become negatively buoyant. This stable stratification would exist even if the core of Venus had the same bulk composition (and thus relative size) as Earth's. In this case, the subsequent thermal evolution of Venus is irrelevant to the prospects for a dynamo. No dynamo would exist even if the core cooled at Earth-like rates. Discovering any crustal remnant magnetism would thus probably disprove this scenario. 

We can extrapolate predictions for exoplanets from these two types of stories about Venus. If tectonic state is the dominant factor, then Venus--like geodynamics should produce Venus--like magnetic histories. That is, a planet with a Venus--like atmosphere (and thus surface) would be less likely to have a long-lived global magnetic field (see Section \ref{sec:VenusThermalHistory}) while modern Venus-like climates might be bad for plate tectonics (see Section \ref{sec:VenusVolatiles}). 
Planetary magnetism could thus serve as a probe of a planet’s tectonic state, which is otherwise difficult to determine by observation. If planet-star distance controls atmospheric properties, then magnetospheres should be rare in the Venus Zone (VZ), but common in the habitable zone (HZ). In contrast, planet-star distance probably does not control the timing of giant impacts during planetary accretion \citep[e.g.][]{Rubie2015,Jacobson2017}. If stochastic events are the dominant factor, then Venus--sized planets in both the VZ and HZ may or may not have magnetospheres. Hence the probability of a global magnetic field would not strongly depend on planet-star distance. Ultimately, exoplanets provide the large sample size necessary to tell us if Venus reflects general principles of planetary evolution, or if Venus trod an evolutionary pathway that is cosmically rare. 

Planetary mass can also affect the prospects for a global magnetic field. The term ``super-Earth" is often used for exoplanets with an Earth-like density but masses up to $\sim$5 Earth-masses and $\sim$1.5 Earth-radii \citep[e.g.][]{Rogers2015,Weiss2014}. However, this terminology may be misleading given the absence of definite facts about the surface of any super-Earth. Any massive planet, especially one in the VZ, could be a ``super-Venus" with a Venus-like atmosphere and hellish surface conditions \cite[e.g.][]{Kane2013}. All else being equal, larger planets are possibly more likely to host dynamos. Larger cores can have higher energy contents \citep[e.g.][]{Driscoll_Olson2011} and, depending on their bulk composition, are still expected to grow solid inner cores that provide a strong power source for a dynamo \cite[e.g.][]{Boujibar2020,Bonati2021,vanSummeren2013}. Simple scaling laws predict that the actual cooling rate of the core would increase with planetary mass faster than the critical value required to drive convection \citep{Blaske_ORourke2021}. Super-Venus (and super-Earth) planets are also likely to have basal magma oceans \citep{Soubiran_Militzer2018} made of liquid silicates that are electrically conductive enough to sustain a dynamo \citep[e.g.][]{Stixrude2020}. Ultimately, a super-Venus could sustain a global magnetic field for much longer than Venus -- meaning that tectonic state and dynamo occurrence might not correlate for massive exoplanets.

\subsection{Atmospheric escape and importance of a magnetic field}\label{sec:VenusMagFieldEscape}

\graphicspath{ {./mag_figures/} }
Here, we discuss present-day observations of the terrestrial planets in our solar system with a focus on Venus, alongside simulations regarding the influence of a global magnetic field on atmospheric escape and habitability. These hold critical lessons for the longevity of exoplanetary atmospheres since the terrestrial worlds of our solar system hold the ground truth necessary to understand atmospheric evolution in general. 

The lack of a global magnetic field at Venus today might lead one to believe that Venus' atmosphere is very vulnerable to the interaction with the solar wind, and thus to the loss of its atmosphere. The effect of the presence of a global magnetic field on atmospheric evolution via atmospheric escape has long been debated. The consensus was that a global magnetic field is important for protecting the atmosphere from being stripped by the solar wind \citep[e.g.][]{Lundin2007}. However, recent spacecraft visiting the three terrestrial sibling planets, Venus, Earth, and Mars, have provided data to shed some new light on this question. Atmospheric escape rates for the three planets appear relatively similar \citep{Strangeway2010}. This new data is important in order to understand if a global magnetic field is necessary for terrestrial planets and exoplanets to retain their atmosphere despite loss caused by stellar radiation. 

To understand the influence of solar wind on atmospheric evolution, we first have to compare the characteristics of the three planets. One of the major differences between them is that Venus and Mars do not have a global magnetic field, while Earth does. Secondly, the size of Venus and Earth is approximately the same, while the radius of Mars is about half of Venus' and Earth's. As a consequence, the mass of Mars is only a tenth of that of Venus or Earth. Third, while Earth's atmosphere is mainly composed of N$_2$ and O$_2$, Venus' and Mars' main atmospheric constituent is CO$_2$. Fourth, Mars has an atmospheric surface pressure of $\approx$6\,mbar, Earth a comfortable 1\,bar, and Venus a crushing 93\,bar. Fifth, as Venus lies closer to the Sun, it resides in a harsher solar radiation and solar wind environment than Earth and Mars. Thus Venus receives about twice and five times more energy and solar wind particles from our host star than the other two planets. It may already be obvious that the solar wind cannot completely remove an atmosphere from a planet even when a global magnetic field is not present, as Venus has the thickest atmosphere of the three sibling planets.

However, we have no constraints on when Venus lost its magnetic field, nor the strength of any field it might have possessed \citep[e.g.][]{ORourke2018}. Thus far, no crustal remnant field has been detected on Venus, as it has been on Mars \citep{Acuna1999}. The crustal remnant magnetic field on Mars tells us that Mars once had a magnetic field, and constraints on its strength can be approximated, even if it is vigorously debated \cite[e.g.][and references therein]{Langlais2019}. Many studies have asserted that remnant magnetism could not survive within the hot crust of Venus. However, at present-day, the surface is $\sim$100 K below the Curie temperatures of common magnetic carriers such as magnetite and hematite. Therefore, crustal remnant magnetism could possibly have survived for billions of years, down to depths of a few kilometers \citep[e.g.][]{ORourke2019}. A magnetometer survey below the ionosphere on a future mission could conduct the first capable search for crustal magnetization \citep{ORourke2018}.

A planet with a global magnetic field will interact with the solar wind and form a magnetosphere, such as at Earth. A planet without a global magnetic field will instead form an induced magnetosphere from the interaction between the solar wind and the ionosphere \citep{luhmann2004induced}, as at Venus and Mars. The difference is important for understanding how the solar wind can influence the escape rates from a planet, as different types of interactions cause different channels of escape to be important.

At Venus, the main escape channels are ion escape from ion pickup in the solar wind or ion acceleration in the magnetotail (for more details see the review of the main Venusian escape channels for O$^+$ and H$^+$ by \citealt{Lammer2006} and in \cite[][,this issue]{Gillmann2022}.
The O$^+$ ion escape rates at Venus have been estimated at $\sim$ $10^{24}-10^{25} \rm s^{-1}$ \citep{Brace1987,McComas1986,Barabash2007,Fedorov2011,Persson2018,Persson2020,Masunaga2019}. These escape rates were also found to be weakly dependent on the solar wind dynamic pressure and energy flux, but not so much with EUV flux \citep{Edberg2011,Kollmann2016,Masunaga2019,Persson2020}. In addition, extreme space weather, such as an Interplanetary Coronal Mass Ejection (ICME) events, may increase the escape rates by several orders of magnitude \citep[e.g.,][]{Luhmann2007}, for a time.

Mars' ion escape rates show a similar order of magnitude to Venus'. The O$^+$ escape rates lie in the range of 
$10^{24} - 10^{25} \rm s^{-1}$ \citep{Bogdanov1975,Lundin1990,Nilsson2012,Ramstad2015,Brain2015,Dong2017,Nilsson2021,Scherf2021}. In contrast with Venus, the O$^+$ escape rates at Mars were found to be inversely correlated with the solar wind dynamic pressure \citep{Dubinin2017,Ramstad2018}, but have a positive correlation with the EUV flux \citep{Ramstad2015}. Due to the lower gravity at Mars, and thus escape velocity, the ions need less acceleration in order to escape, compared to both Venus and Earth. A large part of escape at Mars is therefore the low energy ion escape, which also has a stronger correlation with upstream solar wind and solar XUV flux compared to their higher energy counterparts \citep{Dubinin2017,Ramstad2017}. The escaping ions of less than 50 eV were shown to contribute  between 35-90\% to the total ion escape \citep{Ramstad2017}. 
However, during space weather events it was shown that the high energy ion escape at Mars can increase as it does for Venus \citep{Edberg2010,Jakosky2015}. Hence even though Venus and Mars have the same type of interaction with the solar wind, the escape rates are not dependent on the same parameters.

Despite its strong global magnetic field, Earth displays escape rates of equal or even higher order of magnitude than both Venus and Mars. Several studies indicate average O$^+$ escape rates in the order of $10^{24} - 10^{26} \rm s^{-1}$ \citep[e.g.,][]{Yau1985,Peterson2001,Andersson2005,Nilsson2012,Slapak2017,Schillings2019}. The O$^+$ escape rates at Earth are closely related to geomagnetic activity, and increase with higher activity \citep[e.g.,][]{Yau1985,Slapak2017}. In addition, \citet{Schillings2019} showed that Earth's O$^+$ escape rate is strongly correlated with the solar wind dynamic pressure, but does not have a strong correlation with EUV flux. 

\begin{figure}
\centering
\includegraphics[width=\textwidth]{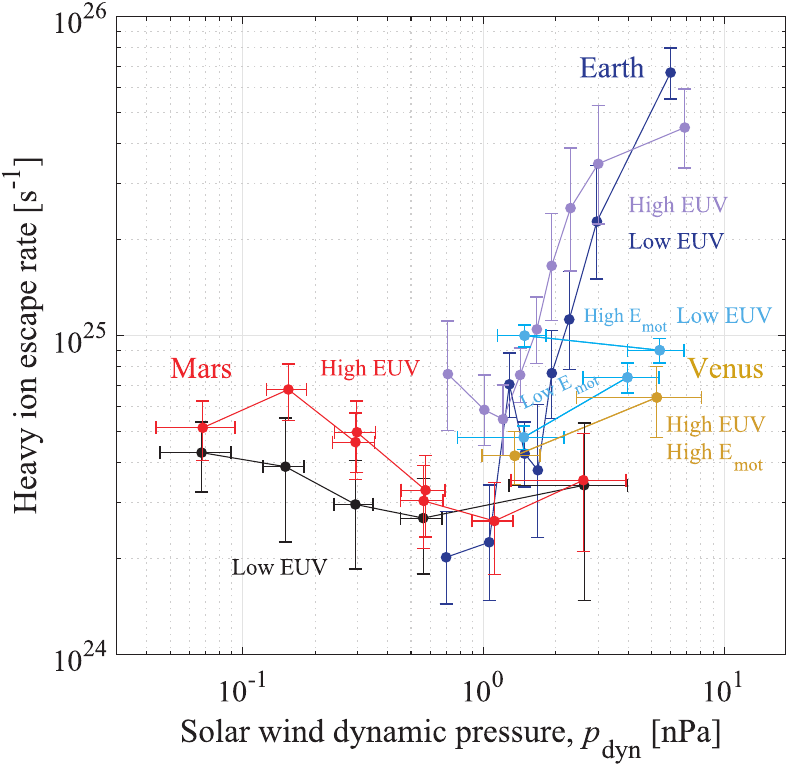}  
\caption{Summary of measured heavy ion escape rates as a function of upstream solar wind dynamic pressure at Venus \citep[blue and yellow,][]{Masunaga2019}, Earth \citep[purple,][]{Schillings2019} and Mars \citep[black and red,][]{Ramstad2018}. Figure adapted from \citet{Ramstad2021}}
\label{fig:esc-comp}
\end{figure}

\begin{figure}
\centering
\includegraphics[width=\textwidth]{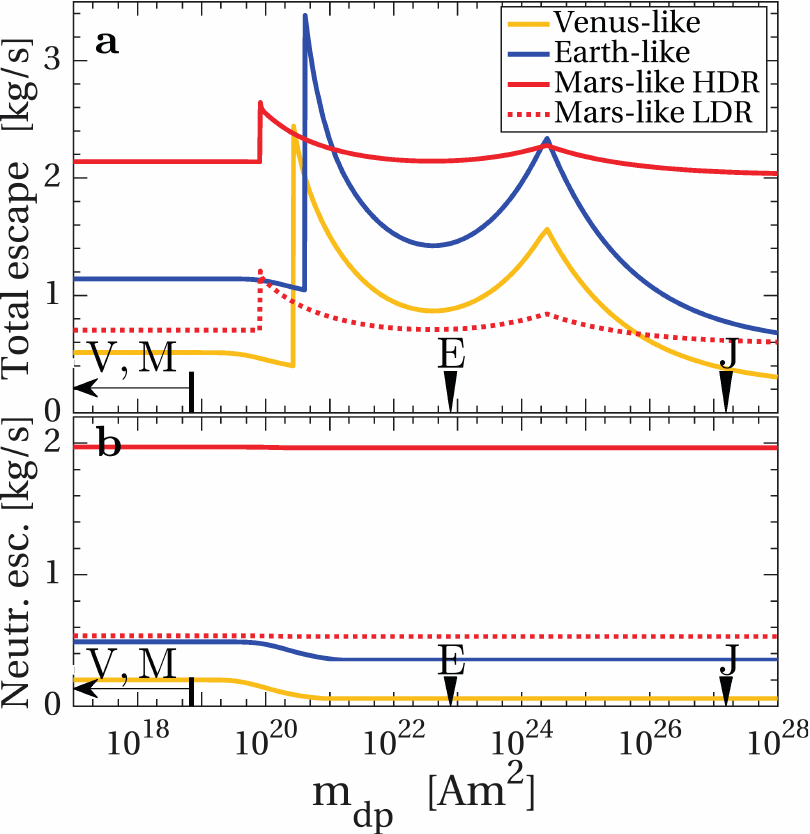} 
\caption{Mass escape from Venus-, Earth- and Mars-like planets, for both neutral and ion (H$^+$ and O$^+$) escape, and how it varies with a change in the dipole magnetic moment of the planet. These are from model computations including seven of the most important escape channels. Today's value of the magnetic moment of Venus (V), Mars (M), Earth (E), and Jupiter (J) is indicated. From \citet{Gunell2018}.}
\label{fig:mag-comp}
\end{figure}

A summary of the results from three studies on the average escape rates at Venus, Earth and Mars is shown in Figure~\ref{fig:esc-comp} as taken from \citet{Ramstad2021}, where the heavy ion escape rates are presented as a function of the solar wind dynamic pressure. As is evident, the escape rates at Earth are higher and more dependent on the changes in the solar wind dynamic pressure than Venus and Mars. \citet{Gunell2018} went into the details on the effect of a global magnetic field on escape by running a set of simulations on how the H$^+$ and O$^+$ escape rates from a Venus-like, an Earth-like and a Mars-like planet would be affected by a change in the dipole magnetic moment of its core. The results of the simulations are shown in Figure~\ref{fig:mag-comp}. They took into account the seven largest escape channels for magnetized and unmagnetized planets. The study gives us a similar picture to the recent measurements shown in  Figure~\ref{fig:esc-comp}: A magnetic field does not always protect the atmosphere, in some cases it can actually increase the escape rates. This conclusion was also supported by global MHD simulations of Venus- and Earth-type exoplanets by \citet{Dong2020}. This means that the global magnetic field is not the only characteristic that determines the escape rate from a planet, there are many other factors to consider.

One important factor to be considered is the composition of a planet's atmosphere, though it tends to be neglected within comparative studies of planetary escape.  While CO$_2$, N$_2$, O$_2$, CO and O heat the upper atmosphere through photoionization by XUV radiation, O$_2$, and O$_3$ through photodissociation by solar UV radiation, and O through exothermic three-body reactions \citep{Kulikov2006}, CO$_2$ molecules  act as an infrared cooler in the thermosphere \citep[e.g.,][]{Roble1989,Roble1995,Mlynczak2010,Cnossen2020}. It emits infrared radiation from the sun back into space, thereby reducing heat within the upper atmosphere. This not only leads to a decline of thermospheric temperature compared to admixtures with less CO$_2$, but also to a decrease of the exobase altitude (see also Gillmann et al. 2022, this issue). IR cooling through CO$_2$ might be the most important of the two effects \citep{Kulikov2006}.

\begin{figure}
\centering
\includegraphics[width=\textwidth]{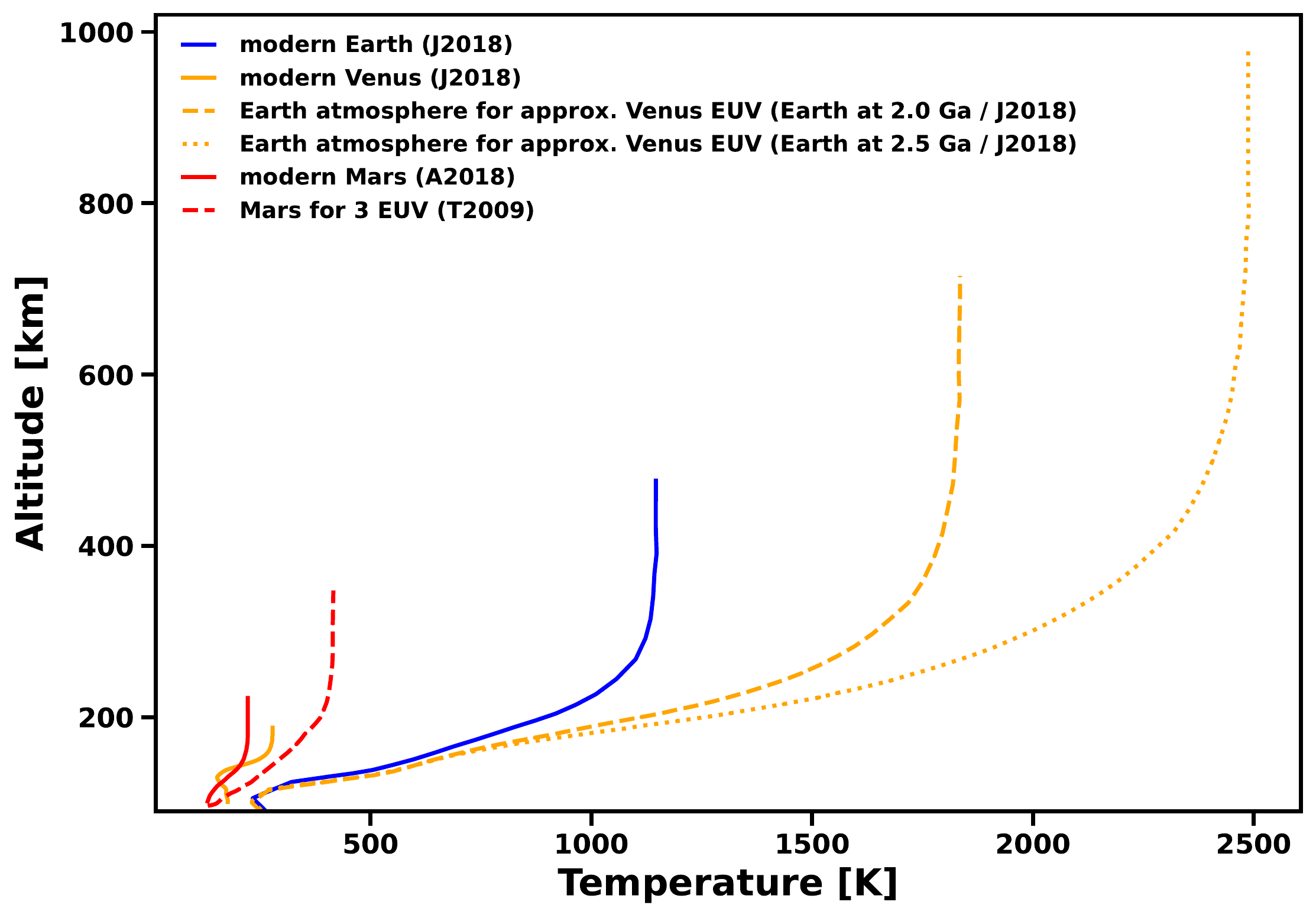} 
\caption{The neutral upper atmosphere profiles for modern Earth \citep{Johnstone2018}, Venus \citep{Johnstone2018}, and Mars \citep{Tian2009}, and for three hypothetical planets \citep{Tian2009,Johnstone2018} that resemble Earth's atmosphere approximately for Venus' EUV flux (dashed and dotted orange lines), and Mars closer to Venus' orbit (the EUV flux at Venus's orbit is about 5 times higher than for Mars, but this plot for 3 EUV is the closest profile available to this value).}
\label{fig:profilesVEM}
\end{figure}

This effect is exemplified through a comparison between the upper atmospheres of Venus and Earth, as can be seen in Figure~\ref{fig:profilesVEM}. Even though Venus receives twice as much energy from our Sun, the altitude of its exobase  ($r_{\rm exo,v} \approx 200$\,km) is less than half that of the Earth ($r_{\rm exo,e} \approx 500$\,km). This is due to the main constituent of the Earth's atmosphere being 78\% N$_2$ and 21\% O$_2$, whereas CO$_2$ only constitutes a minor species (with a mixing ratio of $\approx$0.04\% CO$_2$),
while Venus' atmosphere holds a mixing ratio of about 96\% CO$_2$ and 4\% N$_2$. Mars in turn has a similar atmospheric composition to Venus and a comparable exobase level of $r_{\rm exo,m} \approx 200$\,km. Thus its smaller mass is compensated by an EUV flux that is 5 times less intense than at Venus' orbital distance. In addition to the altitude, CO$_2$ also reduces the average exospheric temperature $T_{\rm exo}$ which varies for neutral particles from about 220\,K and 250\,K at Mars and Venus, respectively, to over 1000\,K at Earth. Both characteristics might affect atmospheric escape.

Figure~\ref{fig:profilesVEM} shows simulated neutral upper atmosphere temperature profiles for present-day Venus \citep{Johnstone2018}, Earth \citep{Johnstone2018}, Mars \citep{Amerstorfer2017}, and three hypothetical planets. The dashed red line \citep{Tian2009} is equivalent to a Martian atmosphere that is irradiated by an EUV flux that is three times as high as at present. For such an increase, exobase level and temperature rise towards $r_{\rm exo} = 415$\,km and $T_{\rm exo} = 350$\,K, respectively. If Mars resided at Venus' orbit, both values would be higher, since the EUV flux at Venus' orbit is about 5 times as high compared to the orbit of Mars. However, this profile is the closest analog to such a planet available in the literature. The dashed and dotted orange lines depict Earth's present-day atmosphere \citep{Johnstone2018} for 2.0 and 2.5 Ga, respectively. This is the approximate time frame at which the EUV flux at Earth's orbit is believed to be about twice as high as at present day (see \citealt{Tu2015}, and Gillmann et al. 2022 this issue), i.e. comparable to the orbital location of Venus. For these two cases, the exobase levels and temperatures for an N$_2$-O$_2$ dominated atmosphere rise towards $r_{\rm exo} = 700$\,km and $T_{\rm exo} = 1800$\,K, and $r_{\rm exo} = 980$\,km and $T_{\rm exo} = 2500$\,K, respectively. If Venus would indeed have such an atmosphere, these levels would be even higher since this planet has a higher equilibrium temperature and about 80\% of the Earth's mass. A nitrogen-oxygen dominated atmosphere around Venus instead of its present-day CO$_2$ atmosphere would, therefore, lead to a significantly different atmospheric structure, thereby illustrating that composition and orbital location indeed matters. But will this also affect the rates of atmospheric escape? Would they cease to be similar if the planets would change place and/or atmospheric composition?

As mentioned earlier, \citet{Gunell2018} derived a formalism to compare atmospheric escape at Venus-, Earth-, and Mars-like planets. Although they did not consider different atmospheric composition, even though this can affect the outcome significantly, as illustrated below. By way of example, these authors \citep[][Equation A.10]{Gunell2018} semi-empirically parameterized the particle loss through ion pickup as,
\begin{equation}\label{eq:pickup}
  Q_{\rm pu,\alpha} = Q_{\rm 0,pu,\alpha} \frac{2 h_a^3 r_{\rm b} h_a^2 r_{\rm b} h_a r_{\rm b}^2}{2 h_a^3 h_a^2 r_{\rm exo} h_a r_{\rm exo}^2} e^{{\frac{\Delta r}{h_{\alpha}}}},
\end{equation}

where $\Delta r = r_{\rm exo} - r_{\rm b}$ is the distance between $r_{\rm exo}$ and the outer boundary layer  $r_{\rm b}$, i.e., either the induced magnetosphere boundary $r_{\rm IMB}$ for an unmagnetized, or the magnetopause standoff distance $r_{\rm sd}$ for a magnetized planet, $h_{\alpha} = (k_{\rm B} T_{\rm exo,\alpha} r_{\rm exo}^2)/G M_{\rm pl} m_{\alpha}$ is the scale height of species $\alpha$, $T_{\rm exo,\alpha}$ is the exospheric temperature of species $\alpha$, $k_{\rm B}$ is the Boltzmann constant, $G$ is the gravitational constant, and $M_{\rm pl}$ is the mass of the planet. The constant $Q_{\rm 0,pu,\alpha}$ is a scaling factor for retrieving today's escape rates in case $r_{\rm exo}$ and $r_{\rm b}$ resemble the present-day values of these planets. As one can see, $r_{\rm exo}$ and $T_{\rm exo}$ are important parameters within $Q_{\rm pu,\alpha}$, and both values are affected by the composition of an atmosphere and the incident EUV flux it receives from its host star. Therefore our hypothetical planets -- Mars with 3 times the present-day EUV flux, and the Venus-like planets with a nitrogen-oxygen dominated atmosphere -- will end up with different values for $Q_{\rm pu,\alpha}$.  

\begin{figure}
\centering
\includegraphics[width=\textwidth]{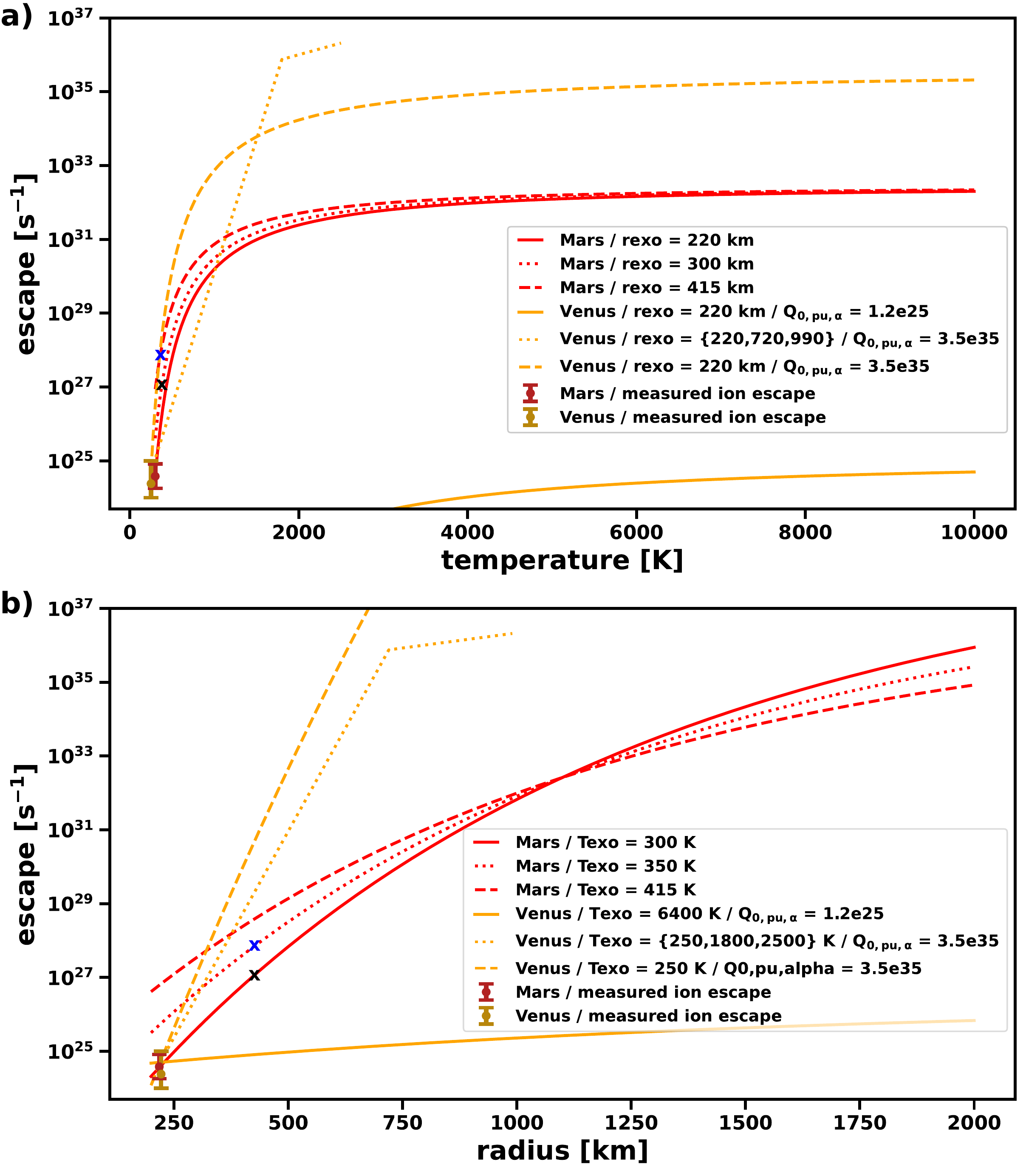} 
\caption{Ion-pickup escape rates of Mars and Venus as calculated with Equation~\ref{eq:pickup} vs. exobase temperature $T_{\rm exo}$ (panel a) and exobase radius $r_{\rm exo}$ (panel b). For Mars, the scaling factor $Q_{\rm 0,pu,\alpha}$  was kept at $2.6 \times 10^{32} \rm s^{-1} = \rm const.$ for all displayed example cases; as one can see, escape rates change significantly for small changes in $T_{\rm exo}$ and $r_{\rm exo}$. For Venus, changes in escape rates are more modest, if the same value for $Q_{\rm 0,pu,\alpha}$ is chosen as in \citet{Gunell2018}. However, if one recalculates $Q_{\rm 0,pu,\alpha}$ by taking into account the exobase temperature of cold oxygen, small changes in $T_{\rm exo}$, again, entail significant changes in escape rates (dashed orange lines). The dotted orange lines illustrate the 3 Venus cases discussed in the main text; here, $T_{\rm exo}$ and $r_{\rm exo}$ were changed simultaneously in both panels. The present-day ion escape rates of Mars and Venus are displayed for comparison; the blue and black crosses are Mars examples discussed in the main text.}
 \label{fig:escEx}
\end{figure}

With this formalism, it is thus in principle possible to directly compare atmospheric loss from Venus, Earth, and Mars with our hypothetical planets. However, it is not straight forward \emph{since we do not know how $r_{\rm IMB}$ scales with the change of exobase level}. Moreover, it turns out that this equation is quite sensitive to the scaling factor $Q_{\rm 0,pu,\alpha}$ and the exobase temperature with which it was derived. This can be seen in Figure~\ref{fig:escEx}, which illustrates how changes in $T_{\rm exo}$ (panel a), $r_{\rm exo}$ (panel b), and $Q_{\rm 0,pu,\alpha}$ (for Venus, both panels -- see below) can affect the outcome of Equation~\ref{eq:pickup} and mostly entail significant changes in ion-pickup escape rates at Mars and Venus. In all of the illustrated cases in Figure~\ref{fig:escEx} $r_{\rm IMB}$ was kept equal to the values employed in \citet{Gunell2018}. Present-day O$^+$ escape rates for Mars and Venus are also shown within this figure; these are displayed for the same values of $T_{\rm exo}$ and $r_{\rm exo}$ as used within \citet{Gunell2018} since there are no specific studies correlating ion escape rates at these planets with different exobase radii and temperatures. A few specific examples of Figure~\ref{fig:escEx} that are related to our hypothetical planets are discussed next.

For Mars, if we keep the scaling factor for oxygen loss at $Q_{\rm 0,pu,\alpha} = 2.6 \times 10^{32} \ \rm s^{-1}$ and insert $r_{\rm exo} = 415$\,km of our hypothetical Martian planet but keep $T_{\rm exo}$ at 300\,K as in \citet{Gunell2018}, the escape rate rises 3--46 times, depending on whether $\Delta r$ or $r_{\rm IMB}$ is kept equal to \citet{Gunell2018} (Figure~\ref{fig:escEx}, black `x' with $r_{\rm IMB}$ kept equal). If we increase the temperature by 50K to $T_{\rm exo} = 350$\,K, then the escape increases even further by about an order of magnitude (Figure~\ref{fig:escEx}, blue `x' with $r_{\rm IMB}$ kept equal).

For our hypothetical Venus-like planets with N$_2$-O$_2$ dominated atmospheres, the change in escape rate is minimal between 1.2 and 4 times for both hypothetical cases and changes in $\Delta r$, if one keeps $T_{\rm exo}$ constant (Figure~\ref{fig:escEx}b, solid orange line). However, \citet{Gunell2018} used the exospheric temperature of hot oxygen to retrieve their scaling factor of $Q_{\rm 0,pu,\alpha} = 1.2 \times 10^{25}\rm s^{-1}$ for oxygen. If we instead scale with the neutral temperature of cold oxygen at the exobase ($\approx$250\,K), which is by far the main oxygen species at the exobase level \citep{Lammer2006}, and retrieve $Q_{\rm 0,pu,\alpha} \approx 10^{35}\rm s^{-1}$, then the loss of oxygen would rise by several orders of magnitude if we insert exobase temperatures of 1800\,K and 2500\,K for our 2.0 and 2.5 Ga cases, respectively (Figure~\ref{fig:escEx}a and b, dotted orange lines). However, this might be above any reasonable escape for such an atmosphere even if it is significantly more expanded than Venus' real atmosphere. 

From an exoplanet perspective this exercise illustrates that it is not trivial to scale the escape and compare different planets with different atmospheric compositions and to draw a definitive conclusion on the importance of intrinsic magnetic fields from the current state of research. Further investigation into atmospheric escape at magnetized and unmagnetized planets is therefore highly warranted. This uncertainty is even more critical if one goes back in time to  higher EUV fluxes than at Venus'  present-day orbit. As already illustrated in Figure~\ref{fig:profilesVEM}, Earth's nitrogen-dominated atmosphere starts to significantly expand for higher EUV fluxes \citep[e.g.,][]{tian2008,Johnstone2018,Johnstone2021}. Crucially, even CO$_2$-dominated atmospheres will start to inflate for fluxes that are about 15 to 20 times higher than at present-day \citep{Tian2009,Johnstone2021}.

Given our present knowledge, it is difficult to estimate how these severely altered conditions (which also apply to young solar-like stars) will affect atmospheric escape, particularly at magnetized planets. \citet{Kislyakova2020} investigated polar escape at Earth for different EUV fluxes ranging back until the Archean eon. They found a significant increase in the polar loss of nitrogen and oxygen within their model from presently $2.1 \times 10^{26}\rm s^{-1}$ and $8.4 \times 10^{24}\rm s^{-1}$ for O$^{+}$ and N$^+$ to $1.6 \times 10^{27}\rm s^{-1}$ and $5.6 \times 10^{26}\rm s^{-1}$ at 2.5 Ga (or 7.6 and 66.7 times more respectively). This increase in escape of O$^+$ is more significant than in the case of unmagnetized Venus, for which it was recently extrapolated back in time by \citet{Persson2020}. However, it is neither well established whether atmospheric escape would have been stronger at Earth without a magnetic field at 2.5 Ga, nor how escape at Venus would have evolved if it had a nitrogen-oxygen dominated atmosphere and/or if it had been ``shielded" by an intrinsic magnetic field. Besides that, it seems probable that a Venus-like exoplanet with an Earth-like atmosphere would show larger escape rates than if it had a CO$_2$-dominated atmosphere, which is important for considering its potential habitability. 
Yet the early Earth atmosphere had very little O$_2$ and a higher pCO$_2$ \citep[e.g.][]{Catling_Zahnle2020} which may have limited atmospheric escape \citep{Lichtenegger2010}. The same possibility exists for early Venus’ atmospheric composition - its evolution would have changed the picture we see today in ways that are difficult to constrain without more information on the planet's distant past.
However, whether an intrinsic magnetic field would diminish the escape remains poorly understood.

From these considerations, one finds that atmospheric composition is likely more important for defining atmospheric loss than the presence of an intrinsic magnetic field. However, even if Earth-like magnetospheres do not shield atmospheres from escape, they can separate particle fluxes according to their energy spectrum so that life forms on a planet's surface are protected from highly energetic primary and secondary solar cosmic rays. There are two sources of cosmic rays, the first originate from high energetic solar events (SCRs), while the second are called galactic cosmic rays (GCRs) that belong to energetic sources in the Milky Way or other galaxies. Upon impact with the Earth's atmosphere, cosmic rays produce showers of secondary particles, some of which reach the surface. SCRs can have global effects on life-forms that enhance mutation rates \citep[][]{Belisheva1995,Belisheva2012,Dar1998,Brack2010}.

Within Earth's magnetospheric cusp area over the Arctic it was found that secondary radiation produced by intense high energy SCR particle showers, like the October 1989 solar proton event \citep{Reeves1992}, caused various biological phenomena associated with DNA lesions on the cellular level \citep{Belisheva1995,Belisheva2012}. These biological effects were detected during experiments with three cellular lines growing in culture during three events of ground level enhancements in the neutron count rate detected and correlated by ground-based neutron monitors, in October 1989 at Srednyi Island, in the White Sea of the Physical Research Institute of the St. Petersburg University, and at the Kola Science Centre of the Russian Academy of Sciences in Apatity, Murmansk region \citep[e.g.,][]{Belisheva2012}. Depending on the planetary magnetic field and atmospheric pressure, cosmic ray particles interact with the atmosphere where they generate secondary highly energetic particles of which some can reach the surface of planets for Earth-like pressure values or lower \citep[e.g.,][]{Shea1995}.

The protection of Earth's surface against secondary high energy solar cosmic ray particles with a surface pressure of $\approx$1\,bar atmosphere amounts to $\approx$1000\,g\,cm$^{-2}$, whereas that of the thin Martian atmosphere with $\leq$10\,mbar only results in $\approx$16\,g\,cm$^{-2}$ \citep[e.g.,][]{Shea1995,Brack2010}. If the planetary atmosphere is dense enough, like that of Venus, these high-energy particles cannot penetrate to the surface. However, the atmospheric region on Venus that may be  favourable for biology is located between and/or near the upper and lower bounds of the three Venusian cloud layers \citep{Cockell1999,Mogul2021,Kotsyurbenko2021} at $\approx$38\,–\,55\,km \citep{Marov1998}, where the atmospheric pressure level is comparable to Earth's. Because Venus is not shielded by an intrinsic magnetosphere like the Earth, high-energy SCR particles will therefore precipitate into its atmosphere and are absorbed around the so-called thermally biological favourable atmospheric layers.

Finally, we point out that smaller magnetic moments that may originate due to tidally locking on terrestrial planets inside the habitable zones of low-mass M and K-type stars, and potentially also due to induced magnetospheres, would provide a weaker protection of planetary surfaces or biologically favourable atmospheric layers against GCRs \citep{Griessmeier2005,Griessmeier2009}. However, in a follow-up study, \citet{Griessmeier2016} point out that for such planets, as well as for unmagnetized bodies, with atmospheric pressures similar or higher than the Earth's, the effects of the increased GCR radiation would be small. For thin atmospheres on the other hand, the shielding from GCRs would be entirely controlled by the magnetosphere, if present. If not, the surface radiation dose cannot be prevented from increasing up to several hundred times the background flux.\newline

\subsection{The Critical Dependence of and on Planetary Thermal History}\label{sec:VenusThermalHistory}

\graphicspath{ {./Thermal_figures/} }

The great divergence between Venus and Earth is critical to understanding potential exoplanetary evolution. Given comparable sizes, masses, and presumably chemical make-up,
Venus is often thought of as the Earth’s twin. As such, one would naturally expect it to exhibit similar patterns of convection, heat loss, and tectonics. Venus, however, is strikingly different in its apparent convective, tectonic, and atmospheric conditions today. These observations lead to a key set of questions: given the broad similarities between Earth and Venus, (1) what led to the dramatic differences between the two planets; and (2) What can the divergence between Venus and Earth tell us about the thermal evolution of exoplanets? With significant attention (in both this chapter and others) devoted to the former, here, we will focus  on the latter. To address this question in some detail, it is important to  outline what we know about the thermal-tectonic regimes and evolution of the Earth. We will then extrapolate this knowledge to the Earth-Venus divergence, and outline potential implications for exoplanets.

The Earth is the only body in the Solar System for which significant information about its thermal, geologic, atmospheric, and tectonic evolution is readily accessible. Consequently, Earth derived data and observations are often used to inform general models of thermal evolution, which are then extrapolated to other bodies in our Solar System, and beyond. However, despite the Earth’s large dataset, our knowledge and understanding of the Earth's thermal evolution remains largely opaque. For instance, while we know Earth is currently within a plate tectonics regime, its initiation and total life of activity are far from certain \citep[e.g.][]{ONeill2007,Debaille2013,Gerya2014,lu2021reviewing}. These uncertain time frames have profound implications for understanding the long-term thermal and surface evolution of the Earth, let alone Venus, and extrasolar planets.   

Critical to this discussion is the notion that the thermal and tectonic state of a planet are intimately connected, and tie into the long-term surface-interior geophysical cycles that influence and control both atmospheric and surface evolution (see Section \ref{sec:VenusVolatiles}; as well as \citealt[][this issue]{Gillmann2022}, \citealt{Phillips2001,Lenardic2008,DriscollBercovici2014,gillmann2014atmosphere,ORourke2018,krissansen2021Venus}). Consequently, a discussion of any one aspect of planetary thermal evolution inherently discusses the other aspects, even if only tacitly. As tectonic states have distinct characteristics, each affects planetary evolution and a planet's thermal state differently. For the purposes of this section, we will briefly outline three main tectonic end-members relative to their thermal implications (definitions of tectonic states are discussed in greater detail in 3.A).

\begin{figure}
\includegraphics[width=\textwidth]{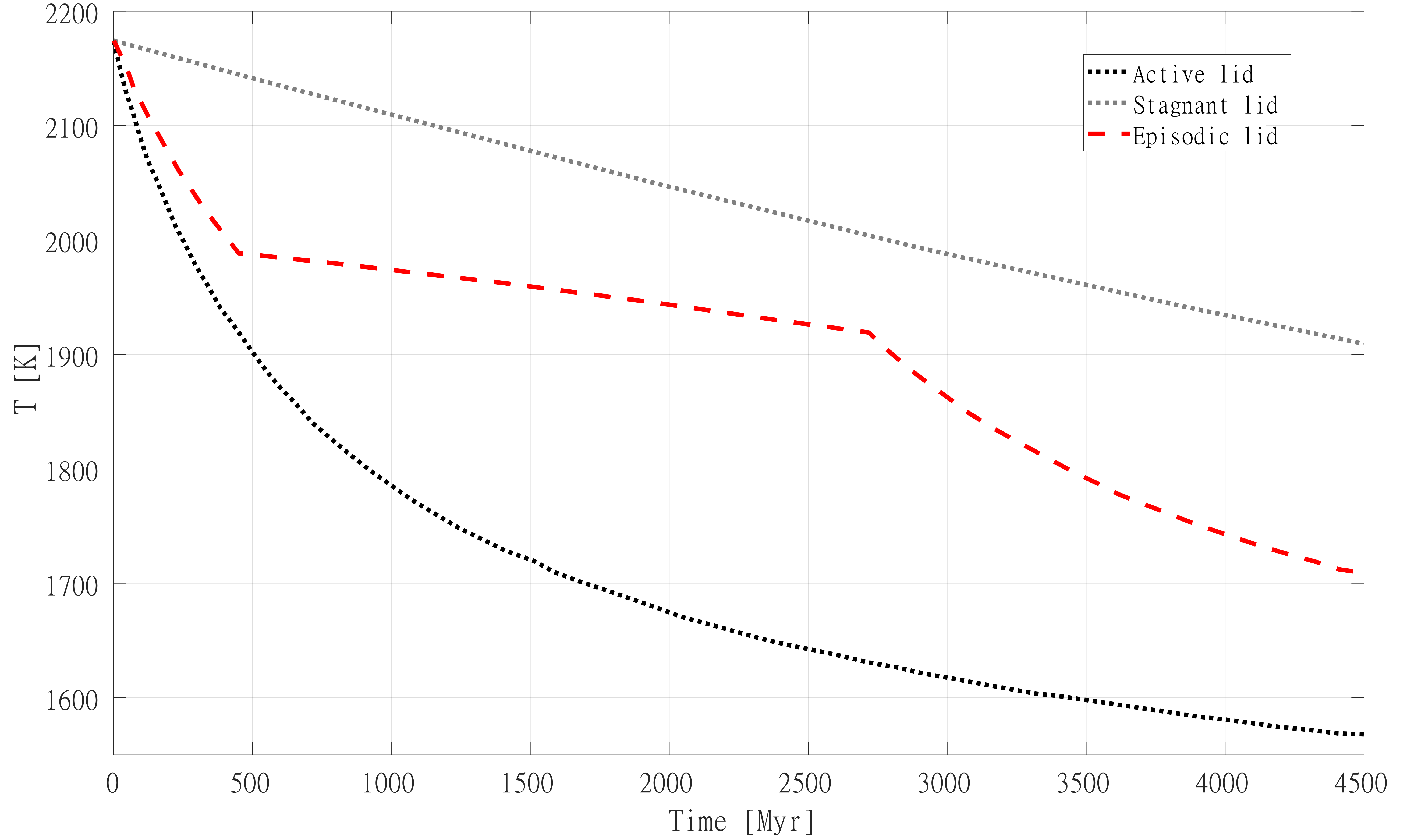}
\caption{Simple thermal history numerical models for an Earth/Venus sized Active lid (black dotted line) and a Stagnant lid (grey dotted line), taken from an identical initial thermal state (here taken as 2174 K) see \cite{Breuer_Moore2007} (and references therein) for a detailed discussion of models. The Episodic lid thermal state is taken from \cite{ONeill2020}, and shows three distinct evolutionary trends: early active episodic, middle quiescent-episodic, and final active lid.
Here T[K] represents the average mantle temperature in Kelvin.
}
\label{fig:thermalhistory}
\end{figure}

Returning to the Earth, we can define plate tectonics as a subset of active (or mobile) lid convection \citep[e.g.][]{Schubert2001}. This mode of tectonics is characterized by the outermost layer of cold and rigid rock participating in the mantle convective cycle. That outer layer is brought back into the interior along with the convective mantle. This leads to the cooling of the interior, a thin lithosphere, and generally efficient heat loss at the surface. In contrast to the mobile lid, the outermost cold and rigid surface layer of the stagnant lid regime resists convective motions \citep[e.g.][]{Schubert2001}. As a consequence, this mode of tectonics has a thicker immobile surface that does not actively participate in mantle convection. The stagnant lid leads to inefficient heat loss and higher internal temperatures when compared to an active lid state.  An additional regime considered is the episodic lid \citep{Moresi_Solmatov1998}, sometimes identified as a transitional regime between active and stagnant lids \citep{Weller2015,Weller_Lenardic2018,Weller_Kiefer2021}. This regime is highly dynamic, characterized by periods of extreme quiescence punctuated with rapid episodes of surface-interior interaction \citep{Armann2012}. In a first order sense, an example of internal temperatures for each regime for an Earth or Venus sized body is indicated in Figure \ref{fig:thermalhistory}. Critical to the discussion of planetary thermal evolution, each of these three states has been suggested to have once operated on the Earth in the past, to varying degrees, though the exact nature and expression of these tectonics, and indeed the thermal state the early Earth exhibited, is vigorously debated \citep[e.g.][]{Condie_Kroner2008,Davies1993,Debaille2013,Calvert1995,ONeill2007,ONeill2015,ONeill2016,Stern2008,Moyen_vanHunen2012,Moore_Webb2013,Gerya2014,ONeill_Debaille2014}. The list of citations is by no means meant to be exhaustive. 

While the geologic record often is ambiguous, and as a consequence, the thermal evolution of the early Earth is passionately debated, it has long been agreed that, as the planet loses heat, the Earth will eventually cease operating in a plate tectonic regime and begin to move into a stagnant-lid regime, similar to observations for current day Mars \citep[e.g.][]{Nimmo_Stevenson2000}. 
While the time frame of this transition remains unclear, a key aspect of planetary tectonics and thermal evolution is highlighted here: the tectonic and thermal state of a planet may change significantly, and perhaps more than once, as the planet evolves. This idea, generally postulated to explain Earth observations, may be extended to other planetary bodies, as has been suggested by studies exploring the convective and tectonic sensitivities to changes in internal mantle temperatures over time, and surface temperature changes through planetary climatic evolution \citep[e.g.][]{ONeill2007,ONeill2016,Lenardic2008,Landuyt_Bercovici2009,Foley2012,Lenardic_Crowley2012,Stein2013,gillmann2014atmosphere,Weller2015,Weller_Lenardic2018}.

Earth and Venus can be seen as planetary end-members (in a bifurcation space). 
For the tectonic/thermal evolution of planets, there exist two main drivers of change: (1) Changes in internal temperatures from changes in heat loss and radiogenic heating rates; and (2) changes in surface temperatures from the long-term climate variations of the planet. 
First, we examine case (1) through the lens of secular cooling (loss of heat with time and depleting internal heat sources). Early in planetary thermal evolution, the internal temperatures are high due to leftover heat from accretion and high levels of radiogenic elements (e.g. Figure \ref{fig:thermalhistory}). From both buoyancy and velocity/stress-scaling arguments \citep[e.g.][and references therein]{Lenardic2021}, these conditions tend to strongly favor early stagnant lid tectonic states \citep{Weller2015,ONeill2016,Weller_Lenardic2018}. 
However, as radiogenic heating, and consequently internal temperatures, decreases with time, this early stagnant state may yield, often through an intermediary episodic state, into an active lid regime. With further heat loss and decrease in radiogenic heating rates, the active lid may ultimately transition once again into a stagnant lid, potentially through an oscillatory episodic state. This stagnant $\rightarrow$ episodic $\rightarrow$ active lid pathway, as suggested for the Earth 
\citep[e.g.][]{ONeill2007,ONeill2016}, can be thought of as the consequence of secular cooling and depletion of radiogenic heating. This then may be thought of as a system state driving force operating on (Earth or Venus sized) planetary bodies, moving the planetary system towards a specific evolutionary path over time, which then may be acted upon by other forces and processes. 

While secular cooling (driver 1) serves to push the planet to an active lid state (and an eventual return to stagnant conditions as more heat is progressively lost), surface temperature changes (driver 2) can profoundly alter the expression of tectonics \citep[e.g.][]{Lenardic2008,Landuyt_Bercovici2009,Foley2012,gillmann2014atmosphere,Weller2015}. For a planet operating in active lid tectonics, an increase in surface temperatures on geologic time scales has been demonstrated to trigger a transition from active lid convection, into a significantly long-lived episodic lid regime \citep{gillmann2014atmosphere}, before eventually settling into stagnant lid behavior \citep{Weller2015,Weller_Kiefer2021}. For an early stagnant lid thermal state, high surface temperatures can prevent the planet from transitioning states entirely. Conversely, a stagnant lid planet with high surface temperature could transition into a mobile lid state, if surface temperature dropped low enough \citep{Lenardic2008,gillmann2014atmosphere}. Therefore, surface temperatures may override the secular driven changes in tectonics for Venus/Earth sized bodies. Alternatively, it could enhance some of its effects, depending on the tectonic/thermal state of a planet at the time of surface temperature change.

\begin{figure}
\includegraphics[width=\textwidth]{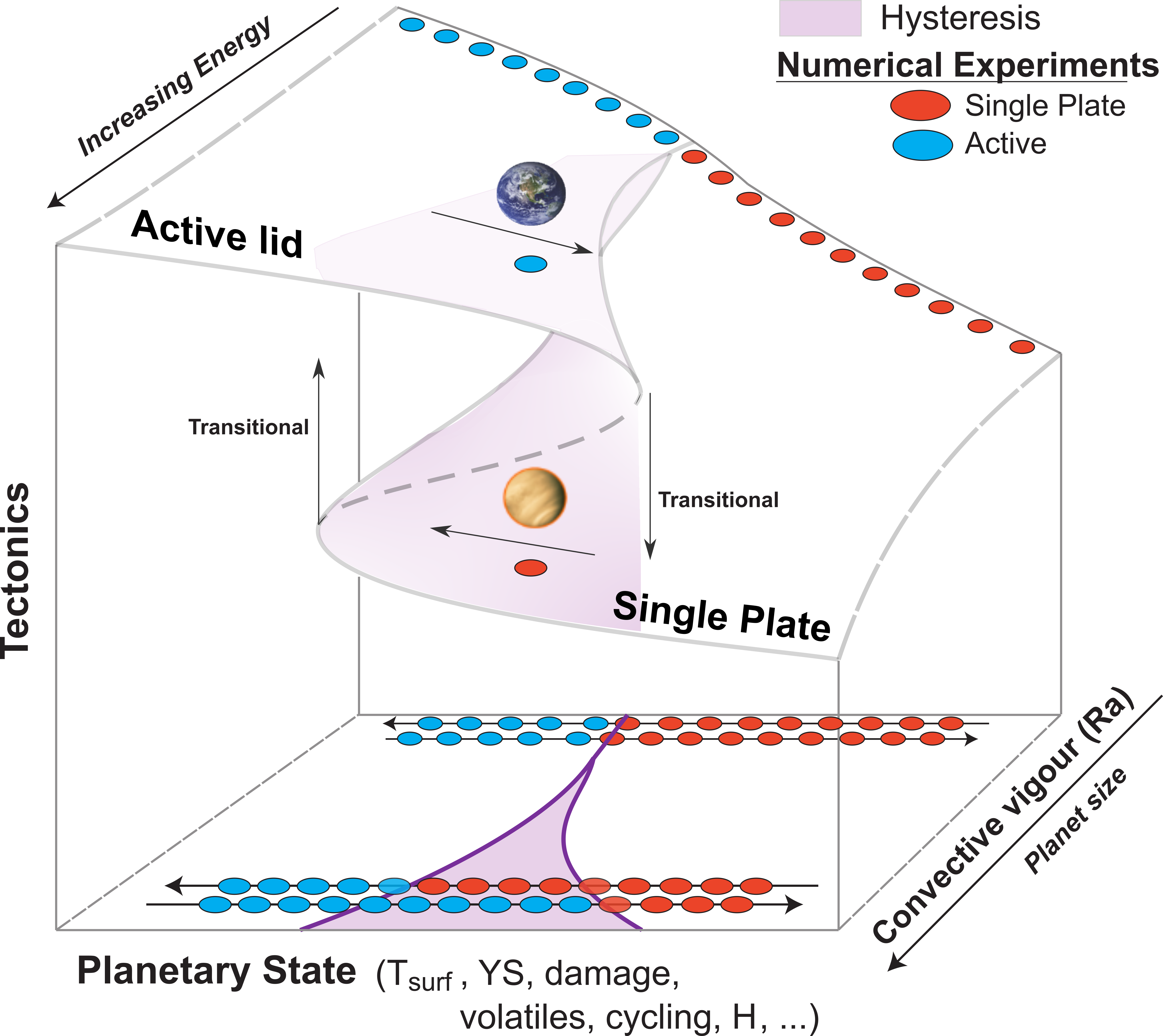}
\vspace{\baselineskip} 
\caption{Modified after \cite{Lenardic_Crowley2012} (Tobias Rolf is credited with an earlier modification of this Figure). Schematic view of bifurcations in planetary tectonics. X-axis denotes changing planetary state variables, for example: Surface temperatures (Tsurf), global yield strength (YS), damage accumulation/healing, volatile abundances and cycling, radiogenic heating rates (H), etc… Convective systems inherently allow for variations in tectonic stability space as a function of increasing convective vigor or energy (Y-axis, background to foreground). For systems with limited energy or low Ra, a single stability point exists (attractor) for a set combination of parameters (e.g. tectonic state has a functional relationship with planetary parameters). For these states, changing parameter paths, or the systems history (denoted by directional increasing/decreasing horizontal arrows with tectonic state indicators: active lid -blue circles, stagnant lid – red circles), has no effect on the final tectonic/thermal state (back projection on the phase space). As complexity increases, multiple attractors effect the stability space for a given set of planetary parameters. Instead on single attractor space (uni-tectonic space), multiple competing attractor wells ensure a path dependence on the final tectonic state. The system allows for rapid changes with parameter variations (direction transition arrows). Multiple solutions exist dependent on the initial conditions and history of the system (hysteresis space, purple shading) as indicated by both mobile and stagnant lid solution viable for the same parameters (foreground). Venus and Earth are plotted as possible endmembers in this hysteresis gap. Putative super-Earth’s/Venus’ would be projected to plot out of the page in ever widening hysteresis space.}
\label{fig:Hysteresis}
\end{figure}

For both early thermal states (hot, young, or enriched in radiogenic/tidal heating sources) and late thermal states (cold, old, or lacking significant radiogenic heating sources), there exists a strong thermal coupling that pushes the planet towards stagnant lid states \citep[e.g.][]{Weller2015,ONeill2016,Weller_Lenardic2018}. However, a significant span of a planet's thermal evolution is controlled by competing and nonlinear forcing, both internal (e.g. heating and temperature) and external (e.g. surface temperature). As a result, the planetary thermal and tectonic state may be predominantly governed by the specific thermal history of the system, allowing stable and unstable active lids, episodic lids, stagnant lids, or all of the above. In fact, nonlinearity within the convective thermal system allows for a hysteresis of states and thermal evolutionary scenarios (Figure \ref{fig:Hysteresis}). Within the hysteresis window, the specific evolutionary history of the system (e.g. the initial conditions, along with the specific thermal evolution) has been shown to play a significant control on the mode of tectonics and thermal state that a planet may operate within. This contrasts with a more traditional view, where a specific set of planetary parameters such as strength of the lithosphere, internal temperature, or surface conditions is directly associated with a specific tectonics/thermal state
\citep{Weller_Lenardic2012,Lenardic_Crowley2012,Weller2015,Weller_Lenardic2018}
(see Figure \ref{fig:Hysteresis} caption).

\vspace{\baselineskip} 
\vspace{\baselineskip}

The hysteresis window is specifically a region of multiple stable tectonic/thermal solutions for otherwise similar planetary bodies. That is, otherwise identical planetary states (e.g. surface temperatures, heating rates, rock strength, volatile contents, etc) can allow for entirely different tectonic and thermal regimes, depending on how the planet evolved toward this state. Interestingly, this window does not seem to be uniform in regard to system complexity or energetics. Figure \ref{fig:Hysteresis} illustrates the hysteresis window conceptually as a function of system energy, or vigor of convection (traditionally considered by the Rayleigh number (Ra) or viscosity contrast). For simple systems with low energy (low convective vigor) there exists a single coupled tectonic-thermal attractor space, or direction of evolution. To put it another way, there exists only one set of stable solutions for any combination of individual planetary states. However, we do not expect planets in general to operate at these low energy/low complexity system states 
\citep{Lenardic_Crowley2012,Weller_Lenardic2012}.
As complexity and the energetics of the system increases (for example Ra and viscosity contrasts), the system is increasingly affected by competing stable tectonic/thermal solutions \citep{Lenardic_Crowley2012,Weller_Lenardic2012}. For conditions expected for real bodies, such as Earth or Venus, the hysteresis space may encompass most reasonable planetary parameters \citep{Weller_Lenardic2018}, 
and consequently the thermal and tectonic evolution of a planet is almost entirely governed by the planet's specific geologic and climatic history. As system complexity and energy increase, as for example for so-called super-Earth’s and super-Venus’, this window may be expected to contain all real solutions. For the foreseeable future, the complexity of such systems make it computationally unfeasible to run in-depth (non-parameterized) numerical simulations to model them.

If we consider a putative proto-Venus/Earth type body, the hysteresis framework offers interesting insight into the coupled thermal tectonic evolution of terrestrial bodies. In this framework, both planetary states are equally possible, and dependent on the specific thermal evolution of each planet. 
In Figure \ref{fig:Hysteresis}, these end-member states are indicated by the Earth evolving along a prior state that allowed active lid convection, whereas Venus’ earlier evolution did not. However, that does not imply that Venus could not have been in an active lid state at some point, or that it fundamentally lacks the capacity to do so. In fact, there exists suggestive, but not unambiguous, evidence that Venus may have operated in some form of active lid mode of tectonics at one time in its past (see \citealt{Rolf2022} this issue for discussion), or that tectonic state may exist as a continuum rather than just simple end-members.

In a general sense, the implications for exoplanets are that there may not exist a preferred tectonic or thermal state for any one planetary variable or type. Instead, the thermal and tectonic state of exoplanets may be much more strongly controlled by the planets’ specific history, a history that we will not be able to sample or observe. As a corollary, this implies that tectonic regime may be vulnerable to change by random events, such as collisions with large impactors \citep{Gillmann2016,ONeill2017}, given they occur at a favourable time to destabilize the current state.
If the planetary tectonic/thermal state of extrasolar planets is non-unique, this suggests that we need to move towards considering tectonic and thermal states in a probability space, as opposed to known variable space (e.g. surface temperature, size, etc). For example, water, if detected in planetary atmospheres, may not be indicative of an active lid state, as has been suggested as the requirement for plate tectonics on Earth \citep{Hubert_Rubey1959,Bird1978}. These results further imply that finding both water and habitable surface conditions would not be an indicator of the tectonic or thermal state of a planet, nor its geologic and climatic history. On the other hand, this probability-oriented approach makes the characterization of exoplanets even more critical to bypass the Solar system assumptions that underpin our understanding of planetary evolution.

Planetary evolution in nonlinear space then is highly complex, but finding solutions is not insurmountable. Instead of focusing on key parameters that control tectonic or thermal states, we need to focus on and understand the probabilities of Venus type solutions relative to Earth (or even other) type solutions. If both Venus and Earth operated within an active lid mode of tectonics in our Solar System, then the potential for active lid modes may be common, but the systems could have strong temporal \citep[e.g.][]{ONeill2016}, stochastic \citep[e.g.][]{Weller_Lenardic2018,Weller_Keifer2020}, and reinforcing feedback \citep[e.g.][]{Lenardic2019} dependencies, that interface in extremely complex ways. The existence of the hysteresis window indicates that we need to understand the feedback effects between the evolution of the atmosphere, mantle, and surface tectonics in a more holistic and probabilistic way through suites of ensemble numerical simulations that focus on the interplay of planetary starting conditions, varying physical parameters and the physics they encompass, as well as stochastic fluctuations. 
Within our own Solar System, results from the InSight mission \citep{Banerdt2020} have greatly improved our understanding of the interior structure of Mars \citep[e.g.][]{KnapmeyerEndrun2021,Khan2021,Stahler2021}. Compared to Mars, which is characterized by a stagnant lid regime throughout its thermal history, Venus tectonic evolution might have been significantly different. Though great care must be taken in extrapolating between dissimilar planets (e.g., Mars to Venus), InSight's results demonstrate how geophysical measurements can provide valuable and detailed information about the interior of other planets. This type of data provides us with the ability to compare and contrast the differences in the interiors of terrestrial planets operating in different tectonic regimes.

The initial thermal state of the planet, which is intimately related to its accretion sequence, determines the amount of energy the planet will dissipate over its history and is thus of fundamental importance regarding its entire evolution.
Despite the absence of direct evidence on the Earth and Venus, several heating mechanisms are thought to affect the earliest stages of planetary evolution \citep[for a detailed discussion, see][this issue]{Salvador2022}. The accretion process itself delivers a substantial amount of energy to the growing planets through the accumulation and burial of impact energy \citep[e.g.,][]{Safronov1978, Tonks1993}. Radiogenic heating produced by the decay of short-lived isotopes (in particular $^{26}$Al and $^{60}$Fe) is responsible for substantial melting of early forming and growing planetary embryos, planetesimals, and proto-planets \citep[e.g.,][]{Merk2002, Bhatia2021}.
During the formation of the core, metal-silicate differentiation and metal downwards migration release gravitational energy dissipated by viscous heating which could increase the temperature of an entire Earth-sized planet by almost 2000 K \citep[][]{Tozer1965, Flasar1973}.
Due to the combination of these heat sources, terrestrial planets are generally thought to experience one or several episodes of early and large-scale mantle melting \citep[e.g.,][]{Elkins-Tanton2012}.
Without an atmosphere overlying the molten surface, the heat accumulated can be rapidly radiated to space but melting can be enhanced and sustained in the presence of a primordial atmosphere \citep[e.g.,][]{Hayashi1979, Ikoma2006} providing a thermal blanketing effect.
On early Earth, the hypothetical Moon-forming giant impact is often referred to as being responsible for generating a last and global-scale magma ocean extending throughout the entire mantle \citep[e.g.][]{Benz1986,Canup2004}. From then on, its cooling, solidification, and associated chemical differentiation would then set the stage for the subsequent long-term evolution of the planet. On Venus, the absence of a moon cannot completely discard the likelihood of an early fully molten stage. Indeed, the orbital proximity of Earth and Venus implies similar bulk properties and suggests that they have experienced similar accretion sequences  \citep[e.g.][]{Morbidelli2012, Raymond2021} with similar endowments of radioactive elements so that the aforementioned heating mechanisms and resulting global-scale melting events would likely apply for both planets, although recent work may put some of this into question \citep[e.g.[]{Emsenhuber2021}.
While these energetic processes are inherent to the formation of terrestrial planets, the initial thermal state and the occurrence and timing of large scale melting events on exoplanets are critically related to the timescale of the accretion phase \citep[see][and references therein]{Salvador2022}. While the current orbital configuration might help put constraints on the tidal heating presently affecting an observed solidified exoplanet, inferring their initial thermal state is out of reach. However, observing a substantial number of young exoplanetary systems might help testing and informing planetary formation and early evolution models to draw more statistically robust trends, thus improving our understanding of early planetary pathways and associated thermal states.

Mantle viscosity is one of the most important parameters that controls the cooling behavior of the interior. This in turn affects magmatic and tectonic processes throughout the thermochemical evolution of the planet. The viscosity of silicate materials is strongly temperature and pressure dependent. The dependence of viscosity on temperature is given by the activation energy, which is the energy necessary to create vacancies in the crystal lattice and the barrier that atoms need to overcome in order to migrate into a vacant site. The activation volume describes the pressure dependence of the viscosity and indicates that for higher pressure the energy necessary for the formation of vacancies and the barrier for atom migration increase. While rheological parameters have been measured in laboratory experiments \citep[e.g.,][]{Hirth2003}, uncertainties in their values are large because such experiments need to be extrapolated to the  conditions relevant for planetary interiors. In particular, the effects of the depth dependence of the viscosity has been highly debated for the deep interior of large rocky planets (super-Earths). Some authors suggest an almost isoviscous interior of large super-Earths indicating a fully convecting mantle \citep{Karato2011}, but others indicate that a strong pressure dependence of the viscosity will lead to the formation of a stagnant region in the lower mantle (the so-called CMB lid) \citep{Stamenkovic2011}. 

\begin{figure}
\includegraphics[width=\textwidth]{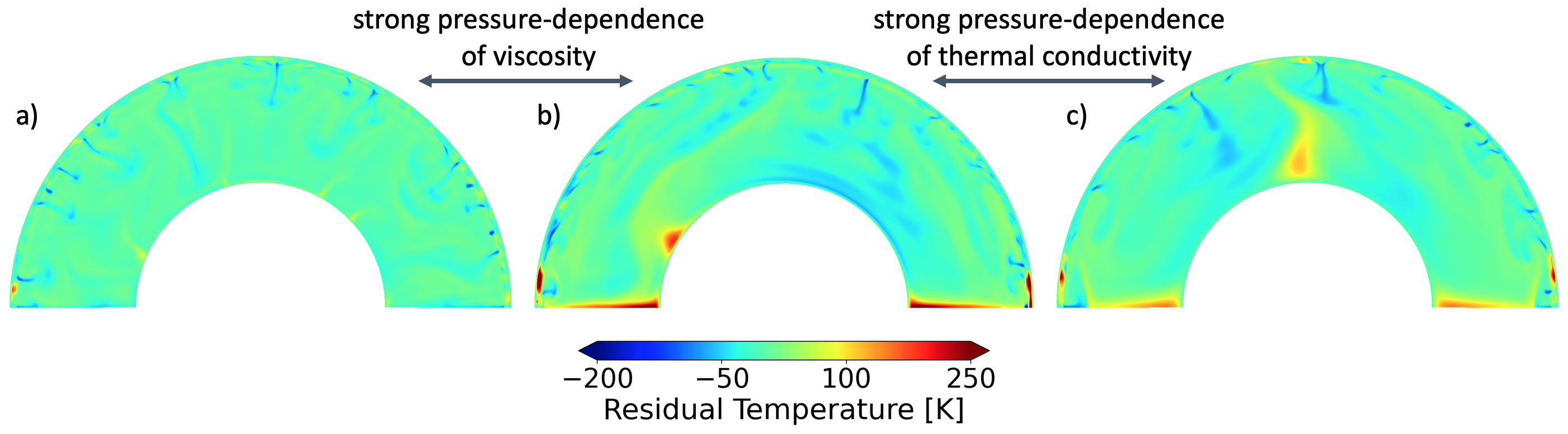}
\vspace{\baselineskip} 
\caption{Effects of pressure dependent parameters on the convection pattern for a Venus-like interior \citep{Hirschberger2020}: a) small pressure dependence of viscosity and thermal conductivity (i.e., viscosity increases with depth by a factor of 32 and thermal conductivity increases with depth by a factor of 1.7); b) strong pressure dependence of the viscosity but weak pressure dependence of thermal conductivity (i.e., viscosity increases with depth by about 4 orders of magnitude and thermal conductivity increases with depth the same as in panel a); c) strong pressure dependence of both viscosity and thermal conductivity (i.e., viscosity increases with depth the same as in panel b and thermal conductivity increases with depth by a factor of about 6).}
\label{fig:mantle_plumes}
\end{figure}

While the pressure inside the mantles of Earth and Venus does not reach the range for which a CMB lid could form, a strong pressure dependence will affect the convection planform, as well as the number and shape of mantle plumes. Mantle convection models show that a strong pressure dependent viscosity will promote fewer and more prominent mantle plumes compared to cases where little or no pressure dependence is applied (Fig. \ref{fig:mantle_plumes}). This in turn may affect the melt production in the interior and the geoid. A strong viscosity increase related to mineral phase transitions, as it is suggested to match the geoid on the Earth, has been found inconsistent with the gravity-topography correlation on Venus \citep{Rolf2018}. This suggests a more gradual increase of the viscosity with depth, possibly indicating a drier upper mantle than on Earth \citep{Rolf2018}. In addition to the viscosity, thermodynamic parameters such as thermal expansivity and thermal conductivity vary with temperature and pressure and can affect the dynamics of the mantle \citep{Tosi2013}. In particular, the increase of thermal conductivity with pressure promotes more diffuse plumes and downwellings thus decreasing the temperature variations in the mantle \citep{Hirschberger2020}. However the strongest effect on convection is expected for the pressure dependence of the viscosity as this increases by several orders of magnitude, compared to an increase by a factor of about 6 for the thermal conductivity \citep{Armann2012}. 

\vspace{\baselineskip} 
\vspace{\baselineskip}

\section{Conclusions}

The terrestrial worlds of our solar system are the benchmarks for exploring the exoplanetary realm of our galaxy. As shown herein there is a tremendous amount of knowledge from solar system objects that can be applied to exoplanetary observations of Venus analogs. Conversely with new ground and space based capabilities coming on-line in the coming decade we will also begin to take lessons from Venus' exoplanetary cousins to learn more about the evolutionary history of Venus and Earth. Yet there is a large imbalance in the knowledge each domain presents us today as reflected in the sizes of the exoplanet versus Venus sections of this chapter. The Venus sections are decidedly larger as one might expect of our nearest planetary neighbor whose atmosphere and surface has been studied intensely with spacecraft and ground based instruments for the past 60+ years, whereas exoplanetary science is still in its infancy. As noted throughout Section 2 Venus studies also benefit tremendously from the study of our home world Earth and our second closest neighbor Mars. For decades planetary scientists have struggled to understand how a possibly early habitable period on both Venus and Mars could result in their present apparently uninhabitable states. If Venus did evolve from an earlier temperate period with surface water reservoirs to it's present hothouse state exactly how did it occur, and what are the key processes involved? We still lack a full understanding of how such a catastrophic event could take place, but there is great anticipation that the study of planets in neighboring stellar systems will help inform our studies of Venus. Yet as shown in Section 1 we are at least two decades away from statistically characterizing the atmospheres of exo-Venus worlds. At the same time we are over a decade away until the data from the newly confirmed Venus missions from ESA and NASA begins to arrive. Even that data will take many years to process and understand, as we see today with the on-going studies of the Magellan Mission radar data from the 1990s \cite[e.g.][]{Byrne2021,Khawja2020,Maclellan2021,Brossier2021,Borrelli2021}.

There are a number of takeaways to consider when looking at how Venus and exoplanetary studies might inform each other in the future as discussed within this chapter. Firstly, lets consider the key role that the early evolution of Venus' magma ocean plays in possibly deciding Venus' long-term H$_2$O budget and the possibility of surface liquid water. In this case exoplanetary observations of planets in the VZ can help us to constrain magma ocean lifetimes around a wide range of stellar hosts, including those explicitly resembling the G-dwarf that is our sun. This involves research programs explicitly looking for solar twins, defined as stellar hosts with chemical compositions or early XUV activity very similar to our sun \citep{Gustafsson2010,Airapetian2021}.
Secondly, why did Earth and Venus take such divergent evolutionary paths when they otherwise appear to be so similar in size, density and possibly chemical composition \citep{Lecuyer2000} in comparison with the other terrestrial planets within the solar system? Examining exoplanets in the VZ may tell us whether Venus ever had temperate surface conditions and whether rotation rate plays a role in stabilizing such conditions as demonstrated in GCM studies \citep{Yang2014,Way2016}. Unfortunately in the near term it could be that a modern Venus-like cloud and haze layer will prevent JWST from resolving atmospheric species that could give clues to exoplanetary atmospheric evolution histories. Clouds in general make observing even major species very challenging with JWST \citep{Fauchez2019,Teinturier2022}, although there may be some opportunities when observing more arid planets with fewer clouds \citep{Ding2022}.
Thirdly, can we discern the longevity of any postulated climate state in Venus' history? For example, if Venus had a temperate period its longevity may be constrained from in-situ observations of the noble gas isotopes as described in \cite[][this issue]{Avice2022} and in \citep{Baines2013}, while exoplanetary worlds in the VZ may also help us to bound the problem. There is an on-going debate as to the timescale of volcanic outgassing required to produce the basaltic plains that cover nearly 80\% of Venus' surface \citep[e.g.][]{Phillips1992,Bullock1993,Herrick1994,Strom1994,BasilevskyHead1996,Bjonnes2012,IvanovHead2013,Kreslavsky2015}. Then there is the nature of the 92 bar CO$_2$ atmosphere in place today. If there was a period of time with a lower atmospheric density (e.g. 1 bar) similar to that achieved by Earth throughout most of its history what mechanism or mechanisms occurred to emplace the present 92 bar atmosphere \citep[e.g.][]{Head2021}? In these last two cases observing a statistically relevant sample of VZ worlds in different evolutionary phases could help us bound the parameter space in ways we may only scarcely comprehend today.

\begin{acknowledgements}
MJW would like to thank Nikolai Piskunov in Uppsala Astrophysics for useful 
discussions regarding the first generation instrumentation on the ELT. We would also like to thank the three anonymous referees who helped us to greatly improve our manuscript.  MJW acknowledges support from the Goddard Space Flight Center  Sellers Exoplanet Environments Collaboration (SEEC) and ROCKE-3D: The evolution of solar system worlds through time, funded by the NASA Planetary and Earth Science Divisions Internal Scientist Funding Model. CG acknowledges the support of Rice University and the CLEVER planets group (itself supported by NASA and part of NExSS). ACP gratefully acknowledges the financial support and endorsement from the DLR Management Board Young Research Group Leader Program and the Executive Board Member for Space Research and Technology. 
\end{acknowledgements}
\bibliographystyle{spbasic}      
\bibliography{references}   
\end{document}